\documentclass[letterpaper,11pt]{article}
\usepackage{amsmath, amssymb}
\usepackage{color}
\usepackage[normalem]{ulem} 
\usepackage{fullpage}
\usepackage[numbers, sort]{natbib}
\usepackage[noend]{algorithmic}
\usepackage{tikz}
\usepackage{pgfplots}
\usepackage{enumerate}
\usepackage{xspace,color}
\usepackage{graphicx}
\usepackage{caption}
\usepackage{subcaption}
\usepackage{enumitem,linegoal}
\usepackage[linesnumbered,ruled,vlined]{algorithm2e}
\usepackage{comment}	
\usepackage{ctable}

\usepackage{mathtools}
\usepackage[flushleft]{threeparttable}
\usepackage{verbatim}
\usepackage{caption}
\usepackage{subcaption}
\usepackage{footnote}

\usepackage{graphicx}
\usepackage{caption}
\usepackage{subcaption}
\usepackage{multicol}

\makesavenoteenv{tabular}
\makesavenoteenv{table}

\usepackage[margin=1in]{geometry}

\definecolor{mygray}{gray}{0.4}
\definecolor{charcoal}{rgb}{0.21, 0.27, 0.31}
\definecolor{coolblack}{rgb}{0.0, 0.18, 0.39}
\definecolor{crimsonglory}{rgb}{0.75, 0.0, 0.2}

 \newtheorem{theorem}{Theorem}[section]
 \newtheorem{lemma}[theorem]{Lemma}
 
 \newtheorem{corollary}[theorem]{Corollary}

 \newtheorem{remark}[theorem]{Remark}
\makeatletter
\def\GrabProofArgument[#1]{ #1: \egroup\ignorespaces}
\def\proof{\noindent\textbf\bgroup Proof%
	\@ifnextchar[{\GrabProofArgument}{. \egroup\ignorespaces}}

\makeatother

\newcommand{\OV}{\textsf{OV}}
\newcommand{\threesum}{\textsf{3-SUM}}
\newcommand{\head}{\mathsf{H}}
\newcommand{\tail}{\mathsf{T}}
\newcommand{\SETH}{\textsf{SETH}}
\newcommand{\SAT}{\textsf{SAT}}
\newcommand{\ttimes}{\star}
\newcommand{\tildorder}{\widetilde O}
\newcommand{\rank}{\mathsf{RK}}
\newcommand{\strassengaussian}{Strassen}

\newcommand{\lefaster}{Le Gall}
\newcommand{\myalpha}{\gamma}
\newcommand{\zwickall}{Zwick}
\newcommand{\bringmanntruly}{Bringmann \textit{et al.}}

\newcommand{\machines}{\mathcal{F}}
\newcommand{\starnumber}{1.145}

\newcommand{\lagal}{Le Gall}
\newcommand{\visible}[1]{{\color{magenta} #1}}
\newcommand*\samethanks[1][\value{footnote}]{\footnotemark[#1]}

\newcounter{proccnt}

\newcommand{\konote}[1]{}

\title{MapReduce Meets Fine-Grained Complexity:\\ MapReduce Algorithms for APSP, Matrix Multiplication, 3-SUM, and Beyond}

\author{
	\and MohammadTaghi HajiAghayi \thanks{University of Maryland, College Park}
	\thanks{Supported in part by NSF CAREER award CCF-1053605,  NSF BIGDATA grant IIS-1546108, NSF AF:Medium grant CCF-1161365,   
		DARPA GRAPHS/AFOSR grant FA9550-12-1-0423, and another DARPA SIMPLEX grant.}
	\and Silvio Lattanzi \thanks{Google}
	\and Saeed Seddighin \samethanks[1] \samethanks[2]
	\and Cliff Stein \thanks{Columbia University, New York, NY 10027, USA} \thanks{Research supported in
		part by NSF grants CCF-1421161 and CCF-1714818. Some research done while visiting
		Google.}
}

\begin{document}
	\newcommand{\ignore}[1]{}
\renewcommand{\theenumi}{(\roman{enumi}).}
\renewcommand{\labelenumi}{\theenumi}
\sloppy

%
%

\date{}

\maketitle

\thispagestyle{empty}

\begin{abstract}
	Distributed processing frameworks, such as MapReduce, Hadoop, and
Spark are popular systems for processing large amounts of data. The
design of efficient algorithms in these frameworks is a challenging
problem, as the systems both require parallelism---since datasets are
so large that multiple machines are necessary---and limit the degree
of parallelism---since the number of machines grows sublinearly in the
size of the data. Although MapReduce is over a dozen years
old~\cite{dean2008mapreduce}, many fundamental problems, such as
Matrix Multiplication, 3-SUM, and All Pairs Shortest Paths, 
 lack efficient MapReduce algorithms. We study these problems
in the MapReduce setting. Our main contribution is to exhibit
smooth trade-offs between the memory available on each machine, and
the total number of machines necessary for each problem. Overall, we
take the memory available to each machine as a parameter, and aim to
minimize the number of rounds and number of machines.

In this paper, we build on the well-known MapReduce theoretical
framework initiated by Karloff, Suri, and Vassilvitskii
~\cite{karloff2010model} and give algorithms for many of these
problems. The key to efficient algorithms in this setting lies in
defining a sublinear number of large (polynomially sized) subproblems,
that can then be solved in parallel. We give strategies for
MapReduce-friendly partitioning, that result in new algorithms for all
of the above problems.  Specifically, we give constant round
algorithms for the Orthogonal Vector (OV) and 3-SUM problems, and
$O(\log n)$-round algorithms for Matrix Multiplication, All Pairs
Shortest Paths (APSP), and Fast Fourier Transform (FFT), among
others. In all of these we exhibit trade-offs between the number of
machines and memory per machine.

\end{abstract}

\maketitle

\section{Introduction}\label{introduction}
During the past decade the amount of user generated data has been
growing at an astonishing rate. For example, even three years ago, 
Facebook's warehouse
stored 300 PB of Hive data, with an incoming daily rate of
about 600
TB~\cite{url1}. Similarly,
Twitter processes over 500 millions tweets per
day~\cite{url2}. As a
consequence, developing parallel and scalable solutions to
efficiently process this wealth of information is now a central
problem in computer science.

\paragraph{MapReduce} 
Distributed processing frameworks such as
MapReduce~\cite{dean2008mapreduce}, Hadoop~\cite{hadoop}, and
Spark~\cite{spark} help address this challenge.  
While differing in details, these frameworks share the same high  level principles. 
The
main advantage of these frameworks is that they: (i) support {\em
  fault-tolerance}, (ii) run on a shared cluster with commodity
hardware, and (iii) provide a simple abstraction to implement new
algorithms. 

In the past
few years several theoretical models have been proposed to enable formal analysis of algorithms in these settings~\cite{feldman2010distributing,karloff2010model,andoni2014parallel,
  goel2012complexity, goodrich2011sorting, kane2010optimal,
  pietracaprina2012space, roughgarden2016, im2017efficient}. Most
notable is the MapReduce model of Karloff, Suri, and Vassilvitskii
\cite{karloff2010model}, which captures the fundamental challenge of
distributing the input data so that no machine ever sees more than a
tiny fraction of it.  So far this model has  received a lot of attention in
the applied community and algorithms for several problems in
clustering, distance measures, submodular optimization, and  query
optimizations have been developed~\cite{bahmani2012scalable, bateni2014distributed,
  im2015brief,ene2015random, kumar2015fast,
  mirzasoleiman2013distributed,beame2013communication,boroujeni2018approximating,hajiaghayi2019massively,boroujeni19}.  A few papers
also considered graph problems such as density, minimum cuts, matchings~\cite{DBLP:conf/spaa/LattanziMSV11,DBLP:conf/soda/AhnGM12,
  ahn2015access, bahmani2012densest, DBLP:conf/kdd/ChierichettiDK14,
  DBLP:conf/kdd/LucierOS15}. Very recently Im, Moseley, and Sun~\cite{im2017efficient} (STOC'17) show how to adapt some dynamic programming algorithms to the MapReduce framework.

MapReduce and its variants are essentially special cases of the Bulk Synchronous Parallel (BSP)
model~\cite{Valiant1990}, but by restricting the allowable parameters,
they better capture what is feasible with modern distributed architectures. 
In the interests of space, we assume familiarity with the basic
MapReduce framework, which can roughly be thought of as alternating
rounds of local computation and global communication. 
MapReduce models have four main parameters to consider: 1) the
number of machines used by the algorithm, 2) the memory available on each
machine, 3) the number of parallel communication rounds, and 4) the
running time of the overall algorithm\footnote{Note that even if we do
  not explicitly restrict the communication in each round, such a
  parameter is bounded by the product of the number of machines and the memory
  used in each machine.}.  The MRC framework popularized by ~\cite{karloff2010model} assumes that 
  for an input of size $n$, the number of machines and the memory per machine is bounded by $O(n^{1 - \epsilon})$ for some $\epsilon >0$, while the number of rounds is polylogarithmic in $n$. 
  
While the MRC framework required the algorithms to take subquadratic
space (since both the number of machines, and the memory per machine
is bounded by $O(n^{1 - \epsilon})$), here we are interested in fine
grained trade-offs between space and the number of machines. To that
end, we will explore the number of machines necessary when memory is
set to $O(n^{\epsilon})$. Obviously, $\Omega(n^{1 - \epsilon})$
machines is a lower bound for all functions that depend on the whole
input, and we strive to get as close to this bound as possible.
Second, we want the algorithm to be {\em work-efficient}, that is,
the total processing time over all machines should be close to the
running time of an efficient sequential algorithm.
 
Work efficiency was an important consideration in many PRAM algorithms
(e.g. \cite{Kruskal1990}), but has received less attention in
MapReduce algorithms.  In the MapReduce setting, in order
to achieve a work efficient-algorithm, we may need to use additional 
total memory.  In fact, there is often a trade-off between how close we are
to a work-efficient algorithm and how much total memory the algorithm is using.  We illustrate this trade of for many classic problems in the MapReduce setting.

Instead of distributing the workload onto different machines,
parallel computing models, such as PRAM, assume that a shared memory
is accessible to several \emph{processors} of the same machine.
Despite the fundamental difference from a practical point of view,
there are many algorithmic similarities between the two models.

For many problems of interest, there are well understood PRAM algorithms.
It is, therefore, natural to ask whether these algorithms can be easily
adapted to new distributed models.
Several papers \cite{karloff2010model, goodrich2011sorting} have shown that it
is possible to simulate the known PRAM algorithms in the MapReduce
model with minimal slow-downs.  But, these simulations have a major drawback -- 
they do not take advantage of the fact that the 
new models are 
\emph{stronger} and allow for \emph{better} algorithms, mainly because
the distributed models allow for free internal 
computation at the machines. Our goal in
designing distributed algorithms is to minimize the total
communication rounds, whereas, in parallel algorithms, the goal is to
minimize the time complexity.

We leverage this ``free'' internal computation in several ways.  In
some cases, it allows us to substantially improve the round complexity
of simulated results, e.g., by achieving constant round (instead of
logarithmic round) solutions. For other cases, it allows us to
improve the overall running time or use fewer processors; and in some
cases, it basically simplifies the known solutions.

\paragraph{Results Overview}

\setcounter{page}{1}
In this paper we propose new MapReduce algorithms for APSP, Matrix Multiplication, 
as well as other problems such as 3-SUM and Orthogonal Vectors whose {\em
  fine-grained complexity} (see~\cite{DBLP:conf/iwpec/Williams15} for
a survey) has been extensively studied and is well understood in the
sequential setting (see
e.g.~\cite{Jia-Wei:1981:ICR:800076.802486,Kerr:1970:EAS:905587}).

We begin with the Orthogonal Vectors and 3-SUM problems a give a few
basic and useful ideas for designing MapReduce algorithms. Next, we
consider more fundamental problems and present our main results. For
matrix multiplication, we show a new technique to parallelize any
matrix multiplication algorithm in the MapReduce framework using the bilinear noncommutative model
introduced by \strassengaussian~\cite{strassen1969gaussian} and
subsequently improved
in~\cite{pan1980new,bini1979n2,schonhage1981partial,romani1982some,coppersmith1982asymptotic,strassen2008asymptotic,coppersmith1990matrix,williams2012multiplying}. 
Interestingly,
while \strassengaussian framework has already been used to develop parallel algorithm in
other computational models~\cite{DBLP:conf/spaa/BallardDHLS12a}, previous approaches
do not extend to the MapReduce framework. Specifically,
we show that given an algorithm based on this method with running time
$O(n^{\omega})$ for matrix multiplication it is possible to obtain a
MapReduce algorithm that, for any $0<\epsilon<1$, uses
$O(n^{\omega(2-\epsilon)/2})$ machines with $O(n^\epsilon)$ memory per
machine. We also extend our approach to non-square matrices.

Next, we present an efficient algorithm for matrix multiplication over
$(\min, +)$, obtaining an efficient
parallel algorithm whose total running time is optimal (i.e. the sum of
the running times over all the parallel instance of the algorithm is
$O(n^3)$).  We then extend it  to new efficient 
MapReduce algorithms for the all pairs shortest paths problem, the
diameter problem and other graph centrality measures. We also show
that using some ideas from~\cite{zwick2002all}, it is possible to improve the
running time of the algorithm if the graph is unweighted or if the
weight of the edges are small.  

\section{Our Results and Techniques}

We present several MapReduce algorithms for fundamental problems.
The novelty of our work is to exhibit smooth trade-offs
between the memory available on each machine, and the total number of
machines necessary. Overall, we take the memory available to each
machine as a parameter, and aim to minimize the number of rounds and
number of machines.

We begin, as a warm-up, by stating our results for the orthogonal
vectors (\OV) problem. This problem is of particular interest to the
fine grained complexity community.
In the \OV\enspace problem,
we are given two lists of vectors, $A$ and $B$, each containing $n$
vectors of size $O(\log n)$, and want to determine whether there
exist two vectors $a \in A$ and $b \in B$, such that $a \cdot b =
0$. Although a quadratic time solution for \OV\enspace is trivial by
iterating over all pairs of vectors and examining whether the inner
product of the vectors is equal to 0,  this algorithm is one of the
fastest algorithms known for \OV\enspace to this date. This solution
is a natural example of an algorithm that can be efficiently
parallelized. Therefore, we begin by presenting a MapReduce algorithm
for \OV.

We show in Section \ref{orthogonalsection} (deferred to Appendix in
the interest of space), that the above algorithm can be implemented in
a single MapReduce round using $O(n^{2(1-\epsilon)})$ machines with
memory $\tildorder(n^{\epsilon})$, for any $0 \leq \epsilon \leq
1$. The idea is to split the lists into sublists of size
$\tildorder(n^\epsilon)$ and assign a machine to every pair of
sublists to find out if the sublists have orthogonal
vectors. Therefore, this algorithm requires $O(n^{2(1-\epsilon)})$
machines. We provide further explanation regarding the MapReduce
details and present pseudocode for both mappers and reducers of this
algorithm in Section \ref{orthogonalsection}.

The idea that the input can be divided into asymptotically smaller
instances, and therefore, problems can be reduced to smaller
subproblems is a promising direction for designing MapReduce
algorithms. However, as we show, this idea does not always lead to the most
efficient algorithms. In Section \ref{3sum}, we study the
\threesum\enspace problem in the MapReduce setting. In this problem,
we are given 3 lists of integer numbers $A$, $B$, and $C$, each
containing $n$ elements. The goal is to determine if there exist $a
\in A$, $b \in B$, and $c \in C$ such that $a+b = c$. Similar to the
solution of \OV, one can divide each of the lists into
$n^{1-\epsilon}$ sublists, each of size $O(n^\epsilon)$. Any combination
of the sublists makes a subtask, and as such, the problem breaks into
$n^{3(1-\epsilon)}$ smaller instances each having an input size of
$O(n^\epsilon)$. Thus, we need $O(n^{3(1-\epsilon)})$ machines to
solve the problem for each combination of the sublists. However,
unlike \OV, \threesum\enspace can be implemented more efficiently in
the MapReduce model. The crux of the argument is that not all
$O(n^{3(1-\epsilon)})$ combinations of sublists need to be examined
for a potential solution. In fact, we show in Section \ref{3sum} that
out of the $O(n^{3(1-\epsilon)})$ combination of sublists, only
$O(n^{2(1-\epsilon)})$ can potentially have a solution. The rest can
be ruled out via an argument on the ranges of the sublists. Therefore,
we can reduce the number of machines needed to solve \threesum\enspace
from $O(n^{3(1-\epsilon)})$ down to $O(n^{2(1-\epsilon)})$. 
The algorithm now needs two rounds, one to determine, on a 
single machine,
 which combinations need to be examined, and a second to 
distribute the subtasks between the machines
and solve the problem.

\vspace{0.2cm}
{\noindent \textbf{Theorem} \ref{3sumtheorem} (restated). \textit{For $\epsilon \geq 1/2$, \threesum\enspace can be solved with a MapReduce algorithm on $O(n^{2(1-\epsilon)})$ machines with memory $O(n^{\epsilon})$ in two MapReduce rounds.}}
\vspace{0.2cm}

Our algorithms for \OV\enspace and \threesum show  how MapReduce
tools can solve classic problems efficiently with less memory per
machine. We now turn to more fundamental problems,
such as matrix
multiplication, graph centrality measures, or shortest
paths. 

Our main contribution is an algorithm for multiplying two $n \times n$
matrices
via MapReduce. Matrix multiplication is one of the most fundamental
and oldest problems in computer science. Many algebraic problems such
as LUP decomposition, the determinant, Gaussian elimination, and
matrix inversion can be reduced to matrix multiplication. 
The trivial algorithm for matrix multiplication takes $O(n^3)$ time, and via a long series
of results the best sequential algorithm currently takes 
$O(n^{\omega^*})$ where $\omega^* =
2.3728$~\cite{le2014powers}.

An important breakthrough in matrix multiplication algorithms was the first improvement
below $O(n^3)$ by Strassen~\cite{strassen1969gaussian} in 1969.  As our algorithms use some of 
the ideas from Strassen, we describe his algorithm here briefly.
\strassengaussian's idea was to show that two $2
\times 2$ matrices can be multiplied only using $7$ integer
multiplications. 
Using recursion, we can think of 
any $n \times n$ matrix as 
four submatrices of size $n/2 \times n/2$, and 
\strassengaussian's observation shows that only $7$ matrix
multiplications of $n/2 \times n/2$ matrices suffice to determine the
solution. Solving the resulting
recursion, we see that 
the total number of integer
multiplications is $O(n^{\log_2 7 \simeq 2.808})$.

Our MapReduce algorithm for matrix multiplication is based on a
similar logic, although instead of
\strassengaussian~\cite{strassen1969gaussian}, we use the latest
decomposition of
\lagal~\cite{le2014powers}. In a single
round, we reduce the problem to smaller instances and divide the
machines between the instances. In the next round, the machines are
evenly divided between the subtasks and each subtask is to multiply
two smaller matrices. Therefore, we again use the same idea to break
the problem into smaller pieces. More generally, in every step, we
divide the matrices into smaller submatrices and distribute the
machines evenly between the smaller instances. We continue this until
the memory of each machine ($O(n^\epsilon))$ is enough to contain all
indices of the matrices, that is, the matrices are of size
$O(n^{\epsilon/2})$. At this point, each machine computes the
multiplication of the given matrices and outputs the solution. We show
in Section \ref{matrixmultiplication} that this can be done in $O(\log
n)$ MapReduce rounds using $O(n^{\omega^*(2-\epsilon)/2})$ machines
and memory $O(n^\epsilon)$ where $\omega^* =
2.3728$~\cite{le2014powers} is the best known upper bound
on the exponent of any $O(n^\alpha)$ algorithm for matrix
multiplication.

\vspace{0.2cm} {\noindent \textbf{Theorem} \ref{squarematrixmult}
  (restated). \textit{For two given $n \times n$ matrices $A$ and $B$
    and $\epsilon > 0$, there exists an $O(\log n)$-round MapReduce
    algorithm to compute $A \times B$ with
    $O(n^{\omega^*(2-\epsilon)/2})$ machines and $O(n^\epsilon)$
    memory on each machine.}
\vspace{0.2cm}

We further extend this result to a MapReduce algorithm for multiplying an $n \times n^x$ matrix into an $n^x \times n$ matrix. Let $\omega^*\langle 1, 1, x\rangle$ denote the smallest $\alpha$ such that one can multiply an $n \times n^x$ matrix into an $n^x \times n$ matrix in time $O(n^\alpha)$. Our algorithm needs $O(n^{\omega^*\langle 1, 1, x\rangle(2-\epsilon)/2})$ machines and $O(n^\epsilon)$ memory on each machine. Similarly, the number of MapReduce rounds of our algorithm is $O(\log n)$.

\vspace{0.2cm} {\noindent \textbf{Theorem} \ref{imbalancedmatrixmult}
  (restated). \textit{ For an $n \times \lceil n^x\rceil $ matrix $A$
    and an $\lceil n^x \rceil \times n$ matrix $B$ and $\epsilon \geq
    0$, there exists an $O(\log n)$-round  MapReduce algorithm to compute $A \times B$
    with $O(n^{\omega^*\langle 1, 1, x\rangle(2-\epsilon)/2})$
    machines and $O(n^\epsilon)$ memory on each machine. }
\vspace{0.2cm}

In Section \ref{plusmin}, we use both Theorems \ref{squarematrixmult} and
\ref{imbalancedmatrixmult} to design similar MapReduce algorithm for
$(\min, +)$ integer matrix multiplication, also known as 
distance multiplication and denoted by
$\ttimes$. 
In this problem, we are given two matrices $A$ and $B$, and
wish to compute a matrix $C$ such that $c_{i,j} = \min_k a_{i,k} +
b_{k,j}$. Again, a trivial $O(n^3)$ solution follows from definition,
however unlike matrix multiplication, the cubic running time of the
naive algorithm has not been improved yet. In fact, many believe that
this problem does not admit any truly subcubic algorithm.
for small ranges,

Our first result for distance multiplication uses a similar approach
as the one we used 
for $\threesum$,  and 
uses
$O(n^{3(1-\epsilon/2)})$ machines and memory $O(n^\epsilon)$ memory on
each machine. This algorithm runs in $1+\lceil
(1-\epsilon/2)/\epsilon\rceil$ MapReduce rounds.

\vspace{0.2cm}
{\noindent \textbf{Theorem} \ref{minplusfirst} (restated). \textit{For any two $n \times n$ matrices $A$ and $B$ and $0 \leq \epsilon \leq 2$, $A \ttimes B$ can be computed with $n^{3(1-\epsilon/2)}$ machines and memory $O(n^{\epsilon})$ in $1+\lceil (1-\epsilon/2)/\epsilon\rceil$ MapReduce rounds.
}}
\vspace{0.2cm}

To prove Theorem \ref{minplusfirst}, we reduce the problem into
smaller instances. However unlike our algorithms for \threesum, we do not assign each
subtask to a single machine. Instead, we assign each subtasks to
several machines and then compute the solution based on the solutions
generated on each of the machines. More precisely, we divide the solution
matrix $C$ into $n^{2(1-\epsilon/2)}$ submatrices, each
of size $n^{\epsilon/2} \times n^{\epsilon/2}$. Notice that for every
$i$ and $j$, we have $c_{i,j} = \min_k a_{i,k} + b_{k,j}$ and
therefore, computing the entries of the solution matrix can be
parallelized by dividing the range of $k$ between the machines. More
precisely, we assign $n^{1-\epsilon/2}$ machines to each submatrix,
with each machine in charge of a range of size $n^\epsilon$ for $k$. Thus, every
machine receives a range and a submatrix along with the corresponding
entries of $A$ and $B$ to that submatrix and outputs the solution. In
the first round, each machine outputs these values for the
corresponding ranges and in the $1+\lceil
(1-\epsilon/2)/\epsilon\rceil$ subsequent rounds, for every $i$ and $j$ we
compute $c_{i,j}$'s based on the generated values in round 1. A
slightly different variant of this algorithm can compute the distance
multiplication of an $n \times n^\epsilon$ matrix into an $n^x \times
n$ matrix with $O(n^{2(1-\epsilon/2)})$ machines.

{\tiny \begin{table}[t]
	\centering
	\begin{tabular}{|c|c|c|}
\hline
problem & \# of machines & \# rounds\\
\hline
\OV\enspace & $O(n^{2(1-\epsilon)})$ & 1\\
(\visible{Theorem \ref{ortho}}) (\visible{\textsf{Work-Efficient}}) && \\
\hline
\threesum\enspace& $O(n^{2(1-\epsilon)})$ & 2\\
(\visible{Theorem \ref{3sumtheorem}}) (\visible{\textsf{Work-Efficient}}) & & \\
\hline
matrix multiplication& $O(n^{\omega^*(2-\epsilon)/2})$ & $O(\log n)$\\
(\visible{Theorem \ref{squarematrixmult}}) (\visible{\textsf{Work-Efficient}})  &  &\\
\hline
LUP decomposition &  & \\
determinant & $\tildorder(n^{\omega^*(2-\epsilon)/2})$ & $O(\log n)$\\
inversion &  & \\
\hline
rectangular matrix multiplication& $O(n^{\omega^*\langle 1,1,x\rangle(2-\epsilon)/2})$ & $O(\log n)$\\
(\visible{Theorem \ref{imbalancedmatrixmult}}) (\visible{\textsf{Work-Efficient}}) & & \\
\hline
distance multiplication& $O(n^{3(1-\epsilon/2)})$ & $O(\epsilon^{-1})$\\
(\visible{Theorem \ref{minplusfirst}})& & \\
\hline
APSP &  & \\
diameter   & &\\
center & $O(n^{3(1-\epsilon/2)})$ & $O(\log n \epsilon^{-1})$\\
negative triangle   &  & \\
(\visible{Corollary \ref{apsp}})  &  & \\
\hline
rectangular distance multiplication& $O(n^{2(1-\epsilon/2)+(x-\epsilon/2)})$ & $O(\epsilon^{-1})$\\
(\visible{Theorem \ref{imbalancedplusmin}}) & &\\
\hline
APSP for unweighted graphs& $O(n^{\machines(\epsilon)})$ & $O(\log ^2 n)$\\
(\visible{Theorem \ref{mainmain}})& &  \\
\hline
	\end{tabular}
\caption{The  number of machines, and the number of MapReduce rounds of our algorithms, using 
$O(n^{\epsilon})$ memory. Here $\omega^*$ is the optimal matrix multiplication exponent; for the definition of $\machines(\epsilon)$ see Section \ref{urizwick}. \textsf{Work-Efficient} means that the total running time of the algorithm is asymptotically equal to the best known sequential algorithm for that problem.}\label{table1}
\end{table}
}

\vspace{0.2cm}
{\noindent \textbf{Theorem} \ref{imbalancedplusmin} (restated). \textit{For $0 < \epsilon \leq 2$ and $x \geq \epsilon$, there exists a MapReduce algorithm that computes $A \ttimes B$ for an $n \times n^x$ matrix $A$ and an $n^x \times n$ matrix $B$. This algorithm runs on $O(n^{2(1-\epsilon/2)+(x-\epsilon/2)})$ machines with memory $O(n^\epsilon)$ and executes in $1 + \lceil (x-\epsilon/2)/\epsilon \rceil$ MapReduce rounds. The total running time of the algorithm over all machines is $O(n^{2+\epsilon})$.}}
\vspace{0.2cm}

{\small 
\begin{table}[t]
	\centering
	\begin{tabular}{|c|c|c|}
\hline
problem & $\max\{\textbf{memory},$ & \\
& $\textbf{\# of machines}\}$  &  rounds\\
\hline
\OV\enspace & $\tildorder(n^{2/3})$ & 1\\
\hline
\threesum\enspace& $O(n^{2/3})$ & 2\\
\hline
matrix multiplication& $O(n^{2\omega^* / (2 + \omega^*)})$ & $O(\log n)$\\
\hline
LUP decomposition &  &  \\
determinant & $O(n^{2\omega^* / (2 + \omega^*)})$ & $O(\log n)$\\
inversion &  &  \\
\hline
distance multiplication& $O(n^{1.2})$ & 2\\
\hline
APSP &  &   \\
diameter  & & \\
center & $O(n^{1.2})$ & $O(\log n)$\\
negative triangle &  &   \\
\hline
APSP for unweighted graphs& $O(n^{1.145})$ & $O(\log ^2 n)$\\
\hline
FFT & $O(n^{0.5})$ & $O(\log n)$\\
\hline
	\end{tabular}
\caption{The complexity of our MapReduce algorithms when the number of machines is equal to the memory of each machine; $\omega^* \approx 2.3728\ldots$ is the best known exponent for matrix multiplication. }\label{table2}
\end{table}
}

A reduction similar to one used by \zwickall\enspace~\cite{zwick2002all}, 
shows that distance multiplication for small ranges can be done
as efficiently as matrix multiplication. We explain this reduction in
more detail in Section \ref{plusmin}. Table \ref{table1} illustrates
the problems that we solve either directly, or via a reduction to
matrix multiplication. Table \ref{table2} shows the performance of our
algorithms when the number of machines is equal to the memory of each
machine.

\subsection{Application to Other Problems}
Matrix multiplication, both over $(\min, +)$ and the standard version
has many applications to other classic problems, and we 
 obtain MapReduce algorithms for 
several of these problems problems. It has been shown that for a weighted graph
$G$ with adjacency matrix $A(G)$ we
have $$\textsf{APSP}(G)=\overbrace{A(G) \ttimes A(G) \ttimes \ldots
  \ttimes A(G)}^{n \text{ times}}$$ where $\textsf{APSP}(G)$ is a
matrix that contains the distance of vertex $j$ from vertex $i$ at
index $(i,j)$. Thus, one can use our algorithm to obtain a MapReduce
algorithm for APSP using $(\min, +)$ matrix multiplication $\lceil
\log n\rceil$ times. This algorithm, then, can be used to determine the
diameter and center of a graph and examine whether a graph contains a
negative cycle. We explain this algorithm in more details in Section \ref{apsp}.

Although APSP can be solved via distance multiplication, we show that
for unweighted graphs (or in general graphs with small weights), the
algorithm can be improved. This improvement is inspired by the work of
\zwickall\enspace~\cite{zwick2002all}. Our approach to obtain this
result in twofold. Recall that APSP can be solved by taking the
adjacency matrix of a graph to the power of $n$ in terms of distance
multiplication, and to compute the $n$th power, we can iteratively apply
distance multiplication $\lceil \log n \rceil$ times. One one hand, we
show that the first few distance multiplications can be run more
efficiently since the entries of the matrices are small,
using the bounded distance multiplication algorithm
instead of the general distance multiplication. On the other hand, the
last distance multiplications can be reduced to rectangular distance
multiplications with fewer than $O(n^2)$ indices. This second  observation is
shown by \zwickall\enspace~\cite{zwick2002all}. Based on these two
ideas, we show in Section \ref{urizwick} that APSP in  unweighted
graphs can be computed more efficiently than in  weighted graphs. 

\vspace{0.2cm}
{\noindent \textbf{Theorem} \ref{mainmain} (restated). \textit{Let $$\machines(\epsilon) = \begin{cases}
		3-5/2\epsilon & \epsilon \leq 0.319\\
		\\
		2(1-\epsilon/2) +&\\
		{\frac{\sqrt{0.07\epsilon^2+1.66\epsilon + -0.43\epsilon+2.92 - 0.89}-0.26\epsilon-1.11}{0.53}-\frac{\epsilon}{2}}, & \text{otherwise}
		\end{cases} $$ 
	 $0 \leq \epsilon \leq 2$ be a real number, and $G$ be a graph with $n$ vertices whose edge weights are in $\{-1,0,1\}$. There exists a MapReduce algorithm to compute APSP of $G$  with $\tildorder(n^{\machines(\epsilon)})$ machines and memory $O(n^\epsilon)$ in $O(\epsilon^{-1} \log^2 n)$ MapReduce rounds.}}
\vspace{0.2cm}

%
%

It has been shown that LUP decomposition, the
determinant, Gaussian elimination, and inversion of a matrix are all
equivalent to matrix multiplication (up to logarithic factors) and thus our algorithms extend to
these problems as well. 


\section{\threesum}\label{3sum}
In \threesum, we are given three lists of integers $A$, $B$, and $C$, each containing up to $n$ numbers. The goal is to find out whether there exist $a_i \in A$, $b_j \in B$, and $c_k \in C$ such that $a_i+b_j = c_k$. The best algorithm for \threesum\enspace on classic computers runs in time $O(n^2)$ and there has not been any substantial improvement for this problem to this date. In fact, many lower bounds are proposed on the running time of several algorithmic problems, based on a conjecture that no algorithm can solve \threesum\enspace in time $O(n^{2-\epsilon})$ for any $\epsilon > 0$~\cite{abboud2013exact}. In this section, we present an efficient MapReduce algorithm for \threesum\enspace that runs in two MapReduce rounds on $O(n^{2(1-\epsilon)})$ machines with memory $O(n^{\epsilon})$. The total running time of the algorithm over all machines is $O(n^2)$. For simplicity, we assume that the numbers of all lists are sorted in non-decreasing order and each list contains exactly $n$ elements. In other words, $A = \{a_1, a_2, \ldots, a_n\}, B = \{b_1, b_2, \ldots, b_n\}, C = \{c_1, c_2, \ldots, c_n\}$ and that for all $i < j$ we have, $a_i < a_j$, $b_i < b_j$, and $c_i < c_j$. 

The classic algorithm for solving \threesum\enspace on one machine is as follows: We iterate over all possible choices of $a_i$ for the first element. For every $a_i$, we create two pointers $p_b$ and $p_c$ initially pointing at the first elements of $B$ and $C$ respectively. Let $v_b$ and $v_c$ denote the values of the pointers at any step of the algorithm. Hence, in the beginning of every iteration, $v_b = b_1$ and $v_c = c_1$ hold. Next, we move the pointers according to the following rule: If $a_i + v_b > v_c$, we push $p_c$ one step forward to point at the next element in the list. Similarly, if $a_i + v_b < v_c$ we change $p_b$ to point to the next element. Otherwise if $a_i + v_b = v_c$, we immediately halt the algorithm and report this triple as an answer to the problem. This way, in every step of the algorithm, we make at most $2n$ iterations and thus the running time of the algorithm is $O(n^2)$. Moreover, since all three lists are sorted initially, we never skip over a potential solution by moving any of the pointers forward. Therefore, if there is any triple $a_i \in A, b_j \in B, c_k \in C$, our algorithm finds it in time $O(n^2)$.

Our MapReduce algorithm for \threesum\enspace is inspired by the above algorithm.  We restrict the memory of each machine to be bounded by $O(n^\epsilon)$ and for simplicity, we assume $1/2 \leq \epsilon \leq 1$. The challenge for the MapReduce setting is that we no longer afford to store all elements of $C$ in a single machine. This is particularly troubling since in order to examine whether a pair of number $(a_i,b_j)$ adds up to some $c_k \in C$, we need to have access to all of the values of the list $C$. To overcome this hardness, our algorithm runs in two MapReduce rounds. We initially divide the $n$ elements of the lists into $n^{1-\epsilon}$ sublists of size $n^{\epsilon}$. For each sublist $X$, we define its head as the smallest number of the list and denote it by $\head(X)$. Similarly, we define the tail of a sublist as the largest number of that list and refer to it by $\tail(X)$. Each sublist contains $n^{\epsilon}$ consecutive numbers of a list and thus for any two sublists $X$ and $Y$ of the same list we have either $\tail(X) \leq \head(Y)$ or $\tail(Y) \leq \head(X)$. In the first MapReduce round of our algorithm, we accumulate all the heads and tails ($O(n^{1-\epsilon})$ many numbers) in a single machine. Based on this information, we determine which triples of the sublists can potentially have a solution. We show the number of such combinations is bounded by $O(n^{2(1-\epsilon)})$. Therefore, the problem boils down to $O(n^{2(1-\epsilon)})$ subproblems of smaller size . In the second MapReduce round of our algorithm, each machine solves a subtask of the problem and finally we report any solution found on any machine as the solution of \threesum. For three sublists $X$, $Y$, and $Z$ of $A$, $B$, and $C$, we say $(X,Y,Z)$ makes a \textit{non-trivial subtask} if $\head(X)+\head(Y) \leq \tail(Z)$ and $\tail(X) + \tail(Y) \geq \head(Z)$. In the following we show how to implement this algorithm with $O(n^{2(1-\epsilon)})$ machines and $O(n^\epsilon)$ memory on each machine.

\begin{theorem}\label{3sumtheorem}
	For $1/2 \leq \epsilon \leq 1$, \threesum\enspace can be solved with a MapReduce algorithm on $O(n^{2(1-\epsilon)})$ machines with memory $O(n^{\epsilon})$ in two MapReduce rounds. The overall running time of our algorithm is $O(n^2)$.
\end{theorem}
\begin{proof}
We first divide each of the lists $A$, $B$, and $C$, into $\lceil n/\ell \rceil = O(n^{1-\epsilon})$ sublists of size $\ell = \lceil n^{\epsilon}\rceil$. Each sublist contains $\ell$ consecutive integers in a sorted list. In the first round of the algorithm we feed the heads and the tails of the sublists to a single machine, and that machine decides how the tasks are distributed among the machines in the second round.

Therefore, in the first round, we only have a single machine working as a reducer. This reducer receives the heads and tails of the sublists and reports all triples of sublists that might potentially contain a solution. Notice that if three sublists do not make a non-trivial subtask then either $\head(X)+\head(Y) > \tail(Z)$ or $\tail(X) + \tail(Y) < \head(Z)$ hold and thus no potential solution can be found in such sublists.

We begin by proving that the number of non-trivial subtasks is $O(n^{2(1-\epsilon)})$.
\begin{lemma}
	The number of non-trivial subtasks is $O(n^{2(1-\epsilon)})$.
\end{lemma}
\begin{proof}
	The crux of the argument is that if $(A_i, B_j, C_k)$ makes a non-trivial subtask, then both $\head(A_i) + \head(B_j) \leq \tail(C_k)$ and $\tail(A_i) + \tail(B_j) \geq \head(C_k)$ hold. Therefore, none of the triples $(A_{i'}, B_{j'}, C_{k'})$ make non-trivial tasks for $i' > i, j' > j, k' < k$ since 
	$$\head(A_{i'}) + \head(B_{j'}) > \tail(A_i) + \tail(B_j) \geq \head(C_k) > \tail(C_{k'}).$$
	
	Now, we show that for any two different non-trivial subtasks $(A_i, B_j, C_k)$ and $(A_{i'}, B_{j'}, C_{k'})$, either $i - j \neq i' - j'$ or $i+j+k \neq i'+j'+k'$. Suppose for the sake of contradiction that both equations hold. We assume w.l.g that $i' > i$ and this implies $j' > j$ since $i - j = i' - j'$ and $k' < k$ since $i+j+k = i' + j' + k'$. This contradicts with the above observation. This shows that if we map every non-trivial subtask $(A_i, B_j, C_k)$ to a pair $(i-j, i+j+k)$, all the corresponding pairs are identical. Notice that both $i-j$ and $i+j+k$ range over intervals of length $O(n^{1-\epsilon})$ and thus the number of non-trivial subtasks cannot be more than $O(n^{1-\epsilon})\cdot O(n^{1-\epsilon}) =  O(n^{2(1-\epsilon)})$.
\end{proof}

Since $\epsilon \geq 1/2$, the memory of a single machine is enough to contain all heads and tails of the sublists. In the first MapReduce round, a reducer identifies all non-trivial subtasks. This can be done in time $O(n^{2(1-\epsilon)})$ on a single machine. To this end,  we first sort the sublists of $A$, $B$, and $C$ based on the values of their heads. It only suffices to find for each pair $(A_i, B_j)$ which sublists of $C$ contain $\head(A_i) + \head(B_j)$ and $\tail(A_i) + \tail(B_j)$. Then we can iterate over all the sublists in between them and report $A_i$, $B_j$, and each of those sublists as a non-trivial subtasks. Note that sorting all pairs of $(A_i,B_j)$ based on $\head(A_i) + \head(B_j)$ can be done in time $O(n^{2(1-\epsilon)})$ since both two sets of sublists are sorted. Therefore, we can determine the sublist of $C$ that contains $\head(A_i) + \head(B_j)$ for all $(A_i, B_j)$ by iterating the sublists of $C$ as well as the sorted pairs of $(A_i, B_j)$. Similarly, we can identify where $\tail(A_i) + \tail(B_j)$ appears for each pair $(A_i, B_j)$ in time $O(n^{2(1-\epsilon)})$. This yields an $O(n^{2(1-\epsilon)})$ time algorithm for identifying all non-trivial subtasks.

In the second MapReduce round of our algorithm, we feed each non-trivial subtask to a single machine and that machine finds out whether there exists a \threesum\enspace solution in that subtasks. Recall that the size of each subtask is $O(n^{\epsilon})$. Therefore, both the number of machines needed for our algorithm and the memory of each machine is $O(n^{\epsilon})$. In addition to this, the running time of each MapReduce phase for every machine is $O(n^{2\epsilon})$ and thus our overall running time is $O(n^2)$.
\end{proof}

\section{Matrix Multiplication}\label{matrixmultiplication}
Matrix multiplication is one of the most fundamental and oldest
problems in computer science. Many algebraic problems such as LUP
decomposition, the determinant, Gaussian elimination, and inversing a
matrix can be reduced to matrix multiplication\footnote{The reductions may incur additional logarithmic factors to the number of machines and the memory of each machine, but these factors are hidden in the $~\tildorder$ notation.}~\cite{pan1985fast}. In addition to this,
matrix multiplication sometimes can be used as black box to solve
combinatorial problems. One example is finding a triangle in an
unweighted graph which can be solved by taking the square of the adjacency matrix of
the graph~\cite{alon1997finding}. Despite the
importance and long-standing of this problem, the computational
complexity of matrix multiplication is not settled yet.

A naive $O(n^3)$ time solution for multiplying two $n \times n$
matrices follows from the definition. 
The first improvement was the surprising result of Strassen, who showed that 
the
multiplication of two $2 \times 2$ can be determined using only 7
integer multiplications, and more generally, that 
the problem of computing the multiplication of
two $n \times n$ matrices reduces to 7 instances of $n/2 \times n/2$
matrix multiplications, plus $O(1)$ additions of $n/2 \times n/2$ matrics, yielding an
$O(n^{\log 7 / \log 2}) \simeq O(n^{2.808})$ algorithm for
matrix multiplication.

Perhaps more important than the improvement on the running time of
matrix multiplication was the general idea of reducing the problem to
small instances in the bilinear noncommutative
model. \strassengaussian~\cite{strassen1969gaussian} showed that 7
multiplications suffice for $2 \times 2$ instances, but any bound on
the number of necessary multiplications for any matrix size can turn
into an algorithm for matrix multiplication. Such a notion is now
known as the rank of a tensor $\langle n,m,k \rangle$ for multiplying
an $n \times m$ matrix by an $m \times k$ matrix and is denoted by
$\rank(\langle n,m,k \rangle)$. Let $\omega$ be the smallest exponent
of $n$ in the running time of any algorithm for computing matrix
multiplication. 
Strassen's algorithm implies $\omega \leq \log \rank{\langle n,n,n \rangle} / \log
n$ and an improvement of 
$\omega \leq 3 \log \rank{\langle
  n,m,k \rangle} / \log nmk$ follows by showing a symmetry on the rank of the
tensors~\cite{lotti1983asymptotic}.
There were then a series of improvements
\cite{pan1980new,bini1979n2,schonhage1981partial,romani1982some,coppersmith1982asymptotic,strassen2008asymptotic,coppersmith1990matrix},
culminating in the result of \lagal\ 
the latest of which shows $\omega \leq
2.3728$~\cite{le2014powers}.   We use $\omega^*$ to denote this bound.
All these bounds are
obtained directly or indirectly by bounding the rank of an $\langle
n,n,n \rangle$ tensor, and thus \lagal\  shows that 
there exists an integer $n_0$ such that $\log \rank(\langle n_0,n_0,n_0\rangle) / \log n_0 \leq \omega^*.$


Moreover, we assume a  decomposition of the tensor $\langle n_0, n_0, n_0 \rangle$ to a corresponding number of products in the bilinear noncommutative model is given since $n_0$ is a constant and one can compute that via exhaustive search. We state our main theorem in terms of $\omega^*$. Indeed, any improvement on the running time of the matrix multiplication based on the bilinear noncommutative model carries over to our setting.
\begin{theorem}\label{squarematrixmult}
For two given $n \times n$ matrices $A$ and $B$ and $\epsilon > 0$, there exists a MapReduce algorithm to compute $A \times B$ with $O(n^{\omega^*(2-\epsilon)/2})$ machines and $O(n^\epsilon)$ memory on each machine. This algorithm runs in $O(\log n)$ rounds.
\end{theorem}
\begin{proof}
The overall idea of the algorithm is to implement the \strassengaussian's idea in a parallel setting. Our algorithm uses $N_p = n^{\omega^*(2-\epsilon)/2}$ machines with memory $N_m = 2n_0^2n^\epsilon$.
Let $l = {n_0}^{\omega^*}$ be the number of terms in a decomposition of the solution matrix. We refer to these terms by $M_1, M_2, \ldots, M_l$ and assume $M_i = \alpha_i \times \beta_i$ where $\alpha_i$ and $\beta_i$ are linear combinations of the indices of $A$ and $B$ respectively. In other words, every entry of matrices $\alpha_i$ and $\beta_i$, is a linear combination of the entries of $A$ and $B$, respectively.
For simplicity, we assume $n$ is divisible by $n_0$ (if not, we add extra 0's to increase the size of the matrix and make its size divisible by $n$). Next, we decompose both matrices into $n_0^2$ submatrices of size $n/n_0$, namely $A_{i,j}$'s and $B_{i,j}$'s. We think of each submatrix $A_{i,j}$ or $B_{i,j}$ as a single entry and compute the product of the two matrices based on the decomposition of ~\cite{le2014powers} in the noncommutative model.

Notice that we only have $(n_0)^2$ elements in the original $n_0 \times n_0$ matrix and thus each $\alpha_i$ and $\beta_i$ is a linear combination of size at most $(n_0)^2$. Therefore, if the number of machines times the memory of each machine is at least $2n_0^2 n^2$, then all $\alpha_i$'s and $\beta_i$'s can be computed in a single round. Thus, we can compute all the variables $\alpha_i$'s and $\beta_i$'s in a single MapReduce round. Via a similar argument, once we compute the values of $M_i$'s for every $1 \leq i \leq l$, then we can in a single round, compute the solution matrix $C$. Therefore, the problem boils down to computing the multiplication of every $\alpha_i \beta_i$ for $1 \leq i \leq l$. We divide the machines evenly between the subproblems and, recursively, compute $M_i$'s in the phase 2 of the algorithm.

In phase 2, for every matrix multiplication of size $n/n_0 \times n/n_0$, we have $N_p/l$ machines with memory $N_m$. We again, use the same method of phase 1 to divide the problem down to $l$ instances of size $n/n_0^2 \times n/n_0^2$ for phase 3. More generally, in phase $i$, for every matrix multiplication of size $n/n_0^{i-1}$ we have $N_p/l^{i-1}$ machines with memory $N_m$. We stop when $i = 1+\log N_p/ \log l$, i.e., we only have a single machine in step $i$. In that case, we compute the matrix multiplication on the only machine dedicated to the subproblem in time $O(n^{\omega^*})$ and report the output. The number of machines and the memory of each machine in each phase is given in Table \ref{zir}.

\begin{table}\centering
\begin{tabular}{|c|c|c|c|}
\hline
Phase \#& Matrix Size & Machines & Memory\\
\hline
Phase 1& $n \times n$ & $N_p$ & $N_m$\\
\hline 
Phase 2& $n/n_0 \times n/n_0$ & $N_p/l$ & $N_m$\\
\hline 
$\vdots$ & $\vdots$ & $\vdots$ & $\vdots$\\
\hline
Phase $i$& $n/n_0^{i-1} \times n/n_0^{i-1}$ & $N_p/l^{i-1}$ & $N_m$\\
\hline
$\vdots$ & $\vdots$ & $\vdots$ & $\vdots$\\
\hline
Phase $\log_l N_p+1$& $n/n_0^{\log_l N_p} \times n/n_0^{\log_l N_p}$ & $1$ & $N_m$\\
\hline
\end{tabular}
\caption{The number of machines, memory of each machine, and size of the matrices in every phase $i$ of the algorithm.}
\label{zir}\end{table}
Notice that in the last phase of the algorithm, the size of the matrices is $n/n_0^{\log_l N_p}$. Moreover, we have $\log l / \log n_0 = \omega^*$, hence, the size of the matrices in the last phase is $n/n_0^{\log_l N_p} = n/n_0^{1/\omega^* \log_{n_0}N_p} =  n/\big(n_0^{\log_{n_0}N_p}\big)^{1/\omega^* } = n/N_p^{1/\omega^*}$. Furthermore, we have $N_p = n^{\omega(2-\epsilon)/2}$ and thus the size of the matrices in the last phase is equal to $n/N_p^{1/\omega^*} = n/n^{(2-\epsilon)/2} = n/n^{1-\epsilon/2} = n^{\epsilon/2}$. Therefore, the memory of the machines in the last phase ($N_m = 2n_0^2n^\epsilon$) suffices to compute the multiplication. Furthermore, the number of machines assigned to each task times the memory of each task is at least $2n_0^2$ times the square of the size of the matrices and thus all linear computations can be done in a single round. Therefore, this algorithm computes the multiplication of two matrices in $O(\log n)$ MapReduce rounds with $N_p$ machines and memory $N_m$.
\end{proof}

Setting $\epsilon = 2\omega^*/(\omega^*+2)$ yields the following corollary.

\begin{corollary}
	For two given $n \times n$ matrices $A$ and $B$, there exists a MapReduce algorithm to compute $A \times B$ with $O(n^{2\omega^*/(\omega^*+2)})$ machines and $O(n^{2\omega^*/(\omega^*+2)})$ memory on each machine. This algorithm runs in $O(\log n)$ rounds.
\end{corollary}
 Note that $2\omega^*/(\omega^*+2) \simeq 1.085$ is very close to 1. The reader can find the complexity of our algorithm in terms of the memory and the number of machines for $n \times n$ matrix multiplication when the number of machines is equal to the memory of each machine for different $\omega$'s in Figure \ref{omegas}.

A closer look at the analysis of Theorem \ref{squarematrixmult} shows that it is not limited to square matrices. For instance, one could show that a similar approach yields to an algorithm for multiplying an $n \times n^x$ matrix into another $n^x \times n$ matrix with fewer than $O(n^{2+x})$ operations. However, in order to use this approach for rectangular matrix multiplication, we need to show a bound on the rank of the $\langle n,n,n^x \rangle$ tensors. To this end, we borrow the result of \lefaster~\cite{le2012faster}.
\begin{theorem}[proven in ~\cite{le2012faster}]\label{sakhtegi}
	Define 	$$\omega\langle 1, 1, x\rangle := \inf_n\{h|\rank(\langle n,n,\lceil n^x\rceil\rangle) = O(n^h)\}.$$
	Then for $\myalpha = 0.30298$ we have 
	$$\omega\langle 1, 1, x\rangle \leq \begin{cases}
	2, & \text{if }x \leq \myalpha \\
	2 + (\omega - 2)(x-\myalpha)/(1-\myalpha), & \text{otherwise.}
	\end{cases}$$
\end{theorem}
Theorem \ref{sakhtegi} allows us to extend Theorem \ref{squarematrixmult} to imbalanced matrix multiplication. In order to multiply an $n \times n^x$ matrix by an $n^x \times n$ matrix for any $0 < x < 1$, , we begin with the following observation: there exists an $n_0$ such that $\rank(\langle n_0,n_0,{n_0}^x\rangle) \leq n_0^{\omega^*\langle 1, x, 1\rangle}$ where $$\omega^*\langle 1, x, 1\rangle  = \omega^*\langle 1, 1, x\rangle
\begin{cases}
2, & \text{if }x \leq \myalpha \\
{2 + \frac{(\omega^* - 2)(x-\myalpha)}{(1-\myalpha)}} \simeq 1.83 + 0.53 x , & \text{otherwise.}
\end{cases}$$ for $\myalpha = 0.30298$~\cite{le2012faster} and $\omega^* = 2.3728$ ~\cite{le2014powers}. This directly follows from Theorem \ref{sakhtegi} and William's bound of $\log \rank(\langle n_0,n_0,n_0\rangle) / \log n_0 \leq \omega^*.$
Let $l = n^{\omega^*\langle 1, x, 1\rangle}$. By
definition, we can formulate the product of an $n_0 \times n_0^x$ by
an $n_0^x \times n_0$ linear combinations of $l$ terms $M_1, M_2,
\ldots, M_l$, where every term $M_i$ is the product of two terms
$\alpha_i$ and $\beta_i$ which are linear combinations of the entries
of each matrix. Similar to what we did in Theorem
\ref{squarematrixmult}, here we use a MapReduce algorithm to compute
this product with several machines of memory $O(n^\epsilon)$.
\begin{theorem}\label{imbalancedmatrixmult}
	For an $n \times \lceil n^x\rceil $ matrix $A$ and an $\lceil n^x \rceil \times n$ matrix $B$ and $\epsilon \geq 0$, there exists a MapReduce algorithm to compute $A \times B$ with $O(n^{\omega^*\langle 1, 1, x\rangle(2-\epsilon)/2})$ machines and $O(n^\epsilon)$ memory on each machine. This algorithm runs in $O(\log n)$ rounds.
\end{theorem}
\begin{proof}
	The proof is similar to Theorem \ref{squarematrixmult}. Let $N_p = n^{\omega^*\langle 1, 1, x\rangle(2-\epsilon)/2}$ be the number of machines and $N_m = 2n_0^2n^\epsilon$ be the memory of each machine. We assume w.l.o.g. that $n$ is divisible by $n_0$ and $\lceil n^x \rceil$ is also divisible by $\lceil {n_0}^x\rceil$. Of course, if that's not the case, one can extend the matrices by adding extra 0's to guarantee these conditions. Also, we assume that each of the two matrices $A$ and $B$ are divided into $n_0\lceil {n_0}^x \rceil$ matrices each having $n/n_0$ rows and $\lceil n^x \rceil  / \lceil {n_0}^x\rceil$ columns. We refer to these matrices by $A_{i,j}$'s and $B_{i,j}$'s.
	
	As stated in the proof of Theorem \ref{squarematrixmult}, our algorithm consists of $O(\log n)$ rounds. In the first round we compute all terms $\alpha_i$'s and $\beta_i$'s for all $1 \leq i \leq l$. This can be done in a single round since we only need to compute  linear combinations of matrices. Then the problem reduces to $l$ different multiplications, each of size $n/n_0 \times \lceil n^x \rceil / \lceil {n_0}^x \rceil$. Once we solve the problem for these matrices, we can, in a single round, compute the solution matrix. In the first round, we have $N_p$ machines with memory $N_m$, and in every phase of recursion the number of machines is divided by $l$ and the size of the problem ($n$) is divided by $n_0$. Therefore, in round $\log_{l}N_p+1$ the size of the problem is $n/{n_0}^{\log_{l}N_p}$. Since $l = {n_0}^{\omega^*\langle 1, 1, x\rangle}$ we have $n/{n_0}^{\log_{l}N_p} = n/N_p^{1/\omega^*\langle 1, 1, x\rangle} = n^{\epsilon/2}$. Moreover, in round $\log_{l}N_p+1$ we only have a single machine with memory $O(n^\epsilon)$ to compute the solution for each subtask. Since the matrices fit in the memory of each machine, we can compute the multiplications in round $\log_{l}N_p+1$ and based on the solutions recursively solve the problem.
\end{proof}

Again, if one wishes to minimize the maximum of the number of machines and the memory of each machine, a bound of $2\omega^*\langle 1,x,1\rangle/(\omega^*\langle 1,x,1\rangle+2)$ can be derived by setting $$\epsilon = 2\omega^*\langle 1,x,1\rangle/(\omega^*\langle 1,x,1\rangle+2).$$
\begin{corollary}[of Theorem \ref{imbalancedmatrixmult}]
	For a given $n \times n^x$ matrix $A$ and an $n^x \times n$ matrix $B$, there exists a MapReduce algorithm to compute $A \times B$ with $O(n^{2\omega^*\langle 1,x,1\rangle/(\omega^*\langle 1,x,1\rangle+2)})$ machines and $O(n^{2\omega^*\langle 1,x,1\rangle/(\omega^*\langle 1,x,1\rangle+2)})$ memory on each machine. This algorithm runs in $O(\log n)$ rounds.
\end{corollary}

Indeed this result improves as the upper bound on $\omega$ improves. Figure \ref{omegas2} shows the exponent ${2\omega^*\langle 1,x,1\rangle/(\omega^*\langle 1,x,1\rangle+2)}$ of the complexity of our algorithm for different $x$'s in the interval $[0,1]$.

\section{Matrix Multiplication over $(\min, +)$}\label{plusmin}
In this section we provide an efficient algorithm for maxtrix multiplication over $(\min, +)$. Given two $n \times n$ matrices $A$ and $B$, our goal is to compute a matrix $C$ such that $c_{i,j} = \min_{k \in [n]}a_{i,k}+b_{k,j}$. Through this paper, we refer to this operation with $\ttimes$. An important observation here is that for any graph $G$ with adjacency matrix $A(G)$, $\overbrace{A(G) \ttimes A(G) \ttimes \ldots \ttimes A(G)}^{n \text{ times}}$ formulates the distance matrix of $G$~\cite{cormen2009introduction}. Therefore, any algorithm for computing $A \ttimes B$ for two $n \times n$ matrices $A$ and $B$ can turn into an algorithm for computing APSP, diameter, and center of a graph with an additional $O(\log n)$ overhead. Thus, all results of this section can be seen as algorithms for computing graph centrality measures. In Section \ref{tikeaval}, we present an algorithm for computing $A \ttimes B$ for two $n \times n$ matrices and show this yields fast algorithms for determining graph centrality measures. Next, we show in Section \ref{tikedovom} that a similar approach gives us an algorithm for imbalanced matrix multiplication over $(\min, +)$. Finally, in Section \ref{tikesevom}, we show that all our results can be improved if the entries of matrices $A$ and $B$ range over a small interval $[-R,R]$.
\subsection{Computing $A \ttimes B$ for $n \times n$ Matrices}\label{tikeaval}
We begin by stating a simple algorithm to compute $A \ttimes B$ in one MapReduce round with $O(n^{2(1-\epsilon)})$ machines and $O(n^{1+\epsilon})$ memory for each machine. We next, show how one can further improve this algorithm by allowing more MapReduce rounds.

The idea is to divide the solution matrix $C$ into $O(n^{1-\epsilon} n^{1-\epsilon}) = O(n^{2(1-\epsilon)})$ submatrices of size $n^{\epsilon} \times  n^{\epsilon}$ and assign the task of computing each submatrix to a separate machine. Each machine then, needs access to the entire rows of $A$ and columns of $B$ ($n^{\epsilon}$ many rows and columns) corresponding its solution matrix and thus its memory is $O(n^{\epsilon} n) = O(n^{1+\epsilon})$. Upon receiving the rows and columns of $A$ and $B$, each machine determines the multiplication of the rows and columns over $(\min, +)$ and reports the output. Therefore, all it takes is a mapper to distribute the rows and columns of the matrices between the machines and $O(n^{2(1+\epsilon)})$ machines to solve the problem for each subtask. The running time of each machine in this case is $O(n^{\epsilon} n^{\epsilon} n) = n^{1+2\epsilon}$. Moreover, the number of machines is $n^{2(1-\epsilon)}$ and thus the overall running time of the algorithm is $O(n^3)$ which is the best known for this problem on classic computers.

Although this seems to be an efficient MapReduce algorithm, we show that this algorithm can be substantially improved to use fewer machines. In the rest of this section, we present an algorithm to compute $A \ttimes B$ with $O(n^{3(1-\epsilon/2)})$ machines and memory $O(n^{\epsilon})$.

\begin{theorem}\label{minplusfirst}
For any two $n \times n$ matrices $A$ and $B$ and $0 \leq \epsilon \leq 2$, $A \ttimes B$ can be computed with $n^{3(1-\epsilon/2)}$ machines and memory $O(n^{\epsilon})$ in $1+\lceil (1-\epsilon/2)/\epsilon\rceil$ MapReduce rounds. Moreover, the total running time of the algorithm is $O((\epsilon^{-1}) n^3)$
\end{theorem}
\begin{proof}
 Our algorithm consists of two stages. The first stage runs in a single MapReduce round. In this round, we divide the solution matrix into $O(n^{1-\epsilon/2} n^{1-\epsilon/2}) = O(n^{2(1-\epsilon/2)})$ matrices of size $n^{\epsilon/2} \times n^{\epsilon/2}$. However, instead of assigning each submatrix to a single machine, this time, we assign the task of computing the solution of each submatrix to $n^{1-\epsilon/2}$ machines. Notice that in order to compute $c_{i,j}$'s, we have to iterate over all $k \in [n]$ and take the minimum of $a_{i,k} + b_{k,j}$ in this range. Therefore, one can divide this job between $n^{1-\epsilon/2}$ machines, by dividing the range of $k$ into $n^{1-\epsilon/2}$ intervals of size $n^{\epsilon/2}$. Each of the machines then, receives a range of size $n^{\epsilon/2}$, a submatrix of the solution, and the corresonding entries of $A$ and $B$ to the solution submatrix and the given range. This makes a total of $O(n^{\epsilon/2} n^{\epsilon/2}) = O(n^{\epsilon})$ matrix entries of $A$ and $B$. Next, each machine finds the solution of its submatrix subject to the range given to it.
The number of solution submatrices is $O(n^{2(1-\epsilon/2)})$. Moreover, the task of solving each submatrix is given to $O(n^{1-\epsilon/2})$ machines and thus the total number of machines used in this round is $O(n^{3(1-\epsilon/2)})$. Furthermore, the memory of each machine in this round is $O(n^{\epsilon/2} n^{\epsilon/2}) = O(n^\epsilon)$, since it only needs to have access to the values of the matrix for its corresponding submatrix of solution and range. Therefore, the momory of each machine is also bounded by $O(n^{\epsilon})$. 

In the first stage we compute $O(n^{1-\epsilon/2})$ values for each entry of the solution. All that is remained is to find the minimum of all these $n^{1-\epsilon/2}$ values for each entry $(i,j)$ of the matrix and report that as $c_{i,j}$. We do this in the second stage of the algorithm. If $\epsilon \geq 2/3$ this can be done in a single MapReduce round as follows: divide the entries of the solution matrix evenly between the machines and feed all related values to each machine. Each machine receive the data associated to $n^2/n^{3(1-\epsilon/2)} = n^{3/2\epsilon-1}$ indices, each having $n^{1-\epsilon/2}$ values generated in the first stage of the algorithm. Notice that the total data given to each machine is $n^{3/2\epsilon-1}n^{1-\epsilon/2} = n^\epsilon$ and thus it fits into the memory of each machine. Next, each machine computes the minimum of all $n^{1-\epsilon/2}$ values generated in the first phase for each index and outputs the corresponding entries of the solution matrix.

The above algorithm fails when $\epsilon < 2/3$. The reason is that no machine has enough memory to contain all $n^{1-\epsilon/2}$ values corresponding to each entry of the solution matrix. However, we can get around this issue by allowing more MapReduce rounds. Since $\epsilon < 2/3$, we have $3(1-\epsilon/2) > 2$ and thus we have more than $n^2$ machines. Therefore, we allocate $n^{3(1-\epsilon/2)}/n^2 = n^{1-3/2\epsilon}$ machines to each entry of the solution matrix. The task of each $n^{1-3/2\epsilon}$ machines is to compute the minimum of all $n^{1-\epsilon/2}$ values for the corresponding entry of the solution matrix. In a single round, we can give $n^\epsilon$ entries to each machine and then compute the minimum of all these numbers in a MapReduce round. This way we can reduce these $n^{1-\epsilon/2}$ numbers to $n^{1-\epsilon/2}/n^\epsilon$ numbers for each entry of the matrix. More generally, in each round we can reduce the size of the data associated to each entry of the matrix by a factor $n^\epsilon$. Thus, we can in $\lceil (1-\epsilon/2)/\epsilon\rceil$ rounds, take the minimum of the data associated to each entry of the solution matrix and report the output.
\end{proof}

As we mentioned earlier, this result carries over to a number of problems regarding graph centrality measures. Included in this list are all pairs shortest paths (APSP), diameter, and center of a graph. Although the number of machines and the memory of each machine remains the same for all these problems, both the running time and the number of MapReduce rounds is multiplied by a factor $O(\log n)$.

\begin{corollary}[of Theorem \ref{minplusfirst}]\label{apsp}
For $0 \leq \epsilon \leq 2$, APSP, diameter, and center of a graph and detecting whether a graph has a negative triangle can be computed with $O(n^{3(1-\epsilon/2)})$ machines with memory $O(n^{\epsilon})$ in $O(\epsilon^{-1} \log n)$ MapReduce rounds. The total running time of the algorithms for these problems is $\tildorder(\epsilon^{-1} n^3)$.
\end{corollary}
\begin{proof}
As we stated before, the APSP matrix of a graph $G$ is equal to the $n$'th power of $A(G)$ (the adjacency matrix of $G$) with respect to $(\min, +)$ matrix multiplication. Of course, this can be done via $\lceil \log n \rceil$ $\ttimes$ operations. Hence, APSP, can be solved by simply using our algorithm for matrix multiplication under $(\min, +)$, $O(\log n)$ times as a blackbox. Negative triangle directly reduces to APSP and thus the same solution works for NT as well~\cite{williams2010subcubic}.  For center and diameter, we first compute the distance matrix of the graph and then in $O(\epsilon^{-1})$ additional MapReduce rounds we find the center or diameter. In these additional MapReduce rounds, for every vertex $v$, we find the closest and furthest vertices to it . This can be done by taking the minimum/maximum of $n$ numbers for each vertex. if $\epsilon \geq 1$, this only requires a single MapReduce round. However, for $\epsilon < 1$, we need $\lceil \epsilon^{-1} \rceil$ MapReduce rounds to take the minimum/maximum of $n$ numbers for each vertex (see the last two paragraphs of Theorem \ref{minplusfirst} for more details).  After this, again we have $n$ numbers indicating the distance of the closest/furthest vertices to each vertex and we wish to find the diameter/center of the graph. This can be again done in $\lceil \epsilon^{-1} \rceil$ rounds by taking the maximum/minimum of these numbers.
\end{proof}

If one wishes to minimize the maximum of the number of machines and the memory of each machine, Theorem \ref{minplusfirst} and Corollary \ref{apsp} yield the following corollaries (by setting $\epsilon$ to $6/5$).
\begin{corollary}[of Theorem \ref{minplusfirst}]
For any two $n \times n$ matrices $A$ and $B$, $A \ttimes B$ can be computed with $O(n^{6/5})$ machines and memory $O(n^{6/5})$ in two MapReduce rounds. Moreover, the total running time of the algorithm is $O(n^3)$.
\end{corollary}

\begin{corollary}[of Corollary \ref{apsp}]
APSP, diameter, and center of a graph can be computed with $O(n^{6/5})$ machines with memory $O(n^{6/5})$ in $O(\log n)$ MapReduce rounds. The total running time of the algorithms for these problems is $\tildorder(n^3)$.
\end{corollary}

\subsection{Computing $A \ttimes B$ for Imbalanced Matrices}\label{tikedovom}
In this section we show that the results of Section \ref{tikeaval} can be extended to imbalance matrices. Let $A$ be an $n \times n^x$ matrix and $B$ be an $n^x \times n$ matrix. In order to compute $C = A \ttimes B$, we can again partition $C$ into smaller $n^{\epsilon/2} \times n^{\epsilon/2}$ matrices and then divide the job of computing each submatrix between many machines. The only difference is that when both matrices are $n \times n$, for each entry $(i,j)$ we need to compute $\min a_{i,k} + b_{k,j}$ for all $k \in [n]$ to determine $c_{i,j}$. Since the memory of each machine is $n^\epsilon$, we have to divide this range between $n^{1-\epsilon/2}$ machines to make sure each interval assigned to each machine is of size $n^{\epsilon/2}$. However, for imbalanced matrix multiplication, because $k$ ranges over $[1, n^x]$, instead of $n^{1-\epsilon/2}$, we need $n^{x-\epsilon/2}$ machines to do this task. Similarly, in the second stage of the algorithm, instead of taking the minimum of $n^{1-\epsilon/2}$ values for each entry of the solution matrix, we need to take the minimum of $n^{x-\epsilon/2}$ values which requires $\lceil (x-\epsilon/2)/\epsilon \rceil$ MapReduce rounds. 

\begin{theorem}\label{imbalancedplusmin}
For $0 < \epsilon \leq 2$ and $x \geq \epsilon$, there exists a MapReduce algorithm that computes $A \ttimes B$ for an $n \times n^x$ matrix $A$ and an $n^x \times n$ matrix $B$. This algorithm runs on $O(n^{2(1-\epsilon/2)+(x-\epsilon/2)})$ machines with memory $O(n^\epsilon)$ and executes in $1 + \lceil (x-\epsilon/2)/\epsilon \rceil$ MapReduce rounds. The total running time of the algorithm over all machines is $O(n^{2+\epsilon})$.
\end{theorem}

By setting $\epsilon = (4+2x)/5$, we can minimize the maximum of the number of machines and the memory of each machine.
\begin{corollary}[of Theorem \ref{imbalancedplusmin}]
For $0 < x \leq 1$, there exists an algorithm to compute $A \ttimes B$ for an $n \times n^x$ matrix $A$ and an $n^x \times n$ matrix $B$. Both the number of machines and the memory of each machine is $O(n^{(4+2x)/5})$ and the algorithm runs in two MapReduce rounds. The total running time of the algorithm over all machines is $O(n^{2+x})$.
\end{corollary}
\subsection{Improvement for Small Range Integers}\label{tikesevom}
In Sections \ref{tikeaval} and \ref{tikedovom}, we presented MapReduce algorithms for $(\min, +)$ matrix multiplication. These algorithms work for the general setting where the indices of the matrices are unbounded integer or real numbers. However, another interesting case to investigate is when the input values are integers in range $[-R, R]$ for small $R$'s. This is particularly interesting since in many problems, the values of the numbers in the matrices come from a small range. For instance, \zwickall~\cite{zwick2002all} showed that a subcubic algorithm for matrix multiplication over $(\min, +)$ for small ranges implies a subcubic algorithm for unweighted APSP. Another example is the work of \bringmanntruly~\cite{bringmann2016truly} wherein an improved algorithm for bounded $(\min, +)$ matrix multiplication is used to obtain a truly subcubic algorithm for language edit distance. In this section, we show that our result for matrix multiplication carries over to the case of bounded $(\min, +)$ matrix multiplication with a small overhead. Later, in Section \ref{urizwick}, we show how to use these results to improve our algorithm for APSP (Theorem \ref{apsp}), for the unweighted case.

In what follows, we show that any algorithm for computing the matrix products over a ring can essentially be turned into an algorithm for computing the $(\min, +)$ matrix multiplication for small ranges. This has been previously proved by \zwickall~\cite{zwick2002all}. Here we just restate the ideas in the MapReduce setting.

\begin{lemma}\label{hadighamgin}
Any MapReduce algorithm for multiplying two matrices with running time $O(n^t)$ on $O(n^x)$ machines and $O(n^y)$ memory running in $z$ MapReduce round implies a MapReduce algorithm for multiplying two matrices of the same size over $(\min, +)$ with $O(n^x)$ machines, memory $\tildorder(R n^y)$, and $z+2$  MapReduce rounds with running time $\tildorder(Rn^t)$, where $R$ is the range of the input values. 
\end{lemma}
\begin{proof}
In order to multiply the two matrices over $(\min, +)$, we only use the matrix multiplication as blackbox. Let the two input matrices be $A$ and $B$, and we wish to compute $C = A \ttimes B$ as the output. We assume  for simplicity that the matrix entries range over $0,R$ but this is w.l.g since any interval of length $R$ can be spanned over $[0,R]$ via a shift.
To multiply the matrices, we construct two matrices $A'$ and $B'$ of the same size as $A$ and $B$ as follows:
\begin{equation*}
a'_{i,j} = (n+1)^{a_{i,j}} \hspace{2cm} b'_{i,j} = (n+1)^{b_{i,j}}
\end{equation*}
where $n$ in the maximum of the rows and columns of the matrices. Notice that the length of each entry of $A'$ and $B'$ is multiplied by a factor $R \log n$. Moreover, it has been shown that addition, subtraction, and multiplication of integers of length $n$ can be done in time $\tildorder(n)$~\cite{aho1974design}. Thus, this only increases the running time and memory by a factor $\tildorder(R)$.  Let $C' = A' \times B'$ be the multiplication of the two matrices with the algorithm described in the theorem. This algorithm uses $O(n^x)$ machines with memory $\tildorder(R n^y)$ and computes $C'$ in $Z$ MapReduce rounds. In what follows, we show that $c_{i,j}$ can be extracted from $c'_{i,j}$. By definition we have $$c'_{i,j} = \sum a'_{i,k}b'_{k,j} = \sum (n+1)^{a'_{i,k}}(n+1)^{b'_{k,j}} = \sum (n+1)^{a_{i,k}+b_{k,j}}.$$
Moreover, since there are at most $n$ terms in the formulation of $c'_{i,j}$, then $c_{i,j} = t$ where $t$ is the smallest number such that $c'_{i,j} \overset{\underset{\mathrm{(n+1)^{t+1}}}{}}{\neq} 0$. Based on this observation, the algorithm to compute $A \ttimes B$ is straightforward. In one MapReduce round, we construct matrices $A'$ and $B'$ from $A$ and $B$ respectively. Then, we multiply $A'$ into $B'$ to obtain $C'$ in $Z$ MapReduce rounds. Finally, in a single MapReduce round, we extract each entry $c_{i,j}$ from $c'_{i,j}$.
\end{proof}

Based on Lemma \ref{hadighamgin}, one can draw the following corollaries from Theorems \ref{squarematrixmult} and \ref{imbalancedmatrixmult}.

\begin{corollary}[of Theorem \ref{squarematrixmult}]\label{hadi}
For two given $n \times n$ matrices $A$ and $B$ with integer entries in the interval $[-R,R]$ and $\epsilon > 0$, there exists a MapReduce algorithm to compute $A \ttimes B$ with $O(n^{\omega^*(2-\epsilon)/2})$ machines and $O(Rn^\epsilon)$ memory on each machine. This algorithm runs in $O(\log n)$ rounds.
\end{corollary}

\begin{corollary}[of Theorem \ref{imbalancedmatrixmult}]\label{hadi2}
Given an $n \times \lceil n^x\rceil $ matrix $A$ and an $\lceil n^x \rceil \times n$ matrix $B$ whose all entries are integers in range $[-R,R]$. For $\epsilon > 0$, there exists a MapReduce algorithm to compute $A \ttimes B$ with $O(n^{\omega^*\langle 1, 1, x\rangle(2-\epsilon)/2})$ machines and $O(Rn^\epsilon)$ memory on each machine. This algorithm runs in $O(\log n)$ rounds.
\end{corollary}

We note that instead of a range $[-R,R]$ for the input values, one can solve the problems for a more general case of range $[-R,R] \cup \{\infty\}$ with essentially the same asymptotic bounds. The idea is that since in matrix multiplication over $(\min, +)$ the maximum value of the solution matrix is bounded by the sum of the maximum values of each of the matrices, one can simply think of $\infty$ as $2R+1$. Therefore, any entry of the solution matrix greater than $2R$ can be replaced by $\infty$.

\begin{remark}
	The bounds of Lemma \ref{hadighamgin} and Corollaries \ref{hadi} and \ref{hadi2} also hold for ranges $[-R,R] \cup \{\infty\}$.
\end{remark}
\section{Acknowledgments}
The authors would like to thank Sergei Vassilvitskii for several fruitful meetings and discussions that led to the  results of the present paper. 


\bibliographystyle{acm}
\bibliography{mapreduce}

\newpage

\appendix
\section*{Appendix}
\section{Improved Algorithm for APSP of Unweighted Graphs}\label{urizwick}
In Section \ref{plusmin}, we present an algorithm for determining the distance matrix of a given weighted graph. This algorithm uses $O(n^\epsilon)$ memory and $O(n^{3(1-\epsilon/2)})$ machines and runs in $O(\log(n))$ MapReduce rounds. Here, we show how to improve this algorithm for the case of unweighted graphs or in general, graphs whose edge weights are in $\{-1,0,1\}$.

Let $G$ be a graph and $A(G)$ be the adjacency matrix of $G$. Moreover, assume that the edge weights of $G$ are either $-1$, $0$, or $1$. Recall that the APSP matrix of $G$ can be computed by taking $A(G)$ to the power of $n$ via $(\min, +)$ multiplication~\cite{cormen2009introduction}. For simplicity of notation, we refer to this  by $A(G)^{\ttimes n}$. 
This observation allows us to compute APSP, using $O(\log n)$ $\ttimes$ operations. Note that since $A(G)$ represents the adjacency matrix of $G$, $A(G)^{\ttimes n} = A(G)^{\ttimes n+i}$ for any positive $i$~\cite{cormen2009introduction}. Therefore, in order to compute $A(G)^{\ttimes n}$, we start with a matrix $S$ initially equal to $A(G)$ and repeat $S \leftarrow S \ttimes S$ $\lceil \log n\rceil$ times. At the beginning of  every step $i$, $S$ is equal to $A(G)^{2^{i-1}}$ and by setting $S \leftarrow S \ttimes S$ we have $S = A(G)^{2^{i}}$. Therefore, after $\lceil \log n \rceil$ $\ttimes$ operations $S$ is the distance matrix of $G$.

The above argument enables us to compute APSP of a graph almost as efficiently as computing $A \ttimes B$ for two $n \times n$ matrices. The only difference here is that our algorithm repeats this multiplication $O(\log n)$ times and thus an additional $O(\log n)$ factor appears in the number of MapReduce rounds and the running time of the algorithm. However, for unweighted graphs, this can be slightly improved. The idea is that if the edge weight are small (either $-1$, $0$, or $1$), then in the first few $\ttimes$ operations, the input values are also small. More precisely, input values are bounded by $2^i$ in every step $i$ of the algorithm. Therefore, if $i$ is small enough, we can instead of the algorithm described in Section \ref{plusmin} (Theorem \ref{minplusfirst}), use the bounded $(\min, +)$ matrix multiplication for $R = O(2^i)$ (Theorem \ref{hadi}). This gives us a bit of improvement on the number of machines needed to compute $S \ttimes S$ for small $i$'s.

Another idea that \zwickall\enspace presented in ~\cite{zwick2002all} deals with the last steps of the algorithm. Notice that for any $1 \leq i \leq n$, $A(G)^{\ttimes i}$ correctly determines the distances of vertices whose size of the shortest path (in terms of the number of edges\footnote{From here on, every time we mention the length of a path, we mean the length in terms of the number of edges.}) is bounded by $i$. Therefore, for large $i$'s, the only difference between $A(G)^{\ttimes i}$ and $A(G)^{\ttimes 2i}$ corresponds to the distance of the vertices whose shortest paths are at least of length $i+1$ (in terms of the number of edges). Now, the crux of the argument is that if for a pair $(u,v)$ of the vertices, we know that the shortest path of $u$ to $v$ has a considerable number of edges, say $k$, then if we randomly sample $n\log n /k$ vertices of the graph, one of the sampled vertices lies on the shortest path of $u$ and $v$ w.h.p. Following this intuition, let us consider a weaker analysis of the algorithm. Previously, we expected that in the $i$'th step of the algorithm, $S = A(G)^{\ttimes 2^i}$ and thus it correctly determines all the shortest paths of length at most $2^i$. In the new analysis, we only demand that in the $i$'th step of the algorithm, $S$ determines the shortest paths of length at most $(3/2)^{i}$. Notice that with this new analysis, we need more steps to make sure all the distances are  determined correctly. Nonetheless, it only multiplies the number of steps by a constant factor. Now, in the beginning of every step $i$, we assume all shortest paths of length at most $(3/2)^{i-1}$ are determined and wish to do so for shortest paths of length at most $(3/2)^i$. Indeed $S \leftarrow S \ttimes S$ satisfies our goal, however, we don't quite need that much computation! 

Notice that any path of length $(3/2)^{i-1} \leq k \leq (3/2)^{i}$ contains at least $(3/2)^i/3$ vertices having a distance of at most $(3/2)^{i-1}$ from both ends (in terms of the number of edges). Updating the distance of $u$ and $v$, from any vertex of this set guarantees a correct solution for $d_{u,v}$ since the distances of these vertices from both ends are already determined correctly in the beginning of the step. Therefore, if we randomly choose $9n/(3/2)^i \log n$ vertices uniformly at random, since one such vertex lies in this set w.h.p, an update with the sampled vertices is enough to compute the shortest paths of size $(3/2)^i$ w.h.p. Let the sampled set be $W$ and $S_1$ be an $n \times |W|$ matrix exactly the same as $S$ except that it only contains the columns corresponding to the vertices of $W$. Similarly, let $S_2$ be a $|W| \times n$ matrix with the same values as $S$, except that $S_2$ only contains rows corresponding to the vertices of $W$. \zwickall~\cite{zwick2002all} argued that $S \leftarrow S_1 \ttimes S_2$ suffices to guarantee that all shortest paths of length $(3/2)^i$ are determined correctly in $S$ w.h.p.

Based on the above arguments, the algorithm that \zwickall~\cite{zwick2002all} proposes for determining the APSP matrix of a graph with small weights ($-1$,$0$, $1$) is as follows:

\begin{algorithm}
	\KwData{A graph $G$ with edge weights in $\{-1,0,1\}$}
	\KwResult{The APSP matrix of the graph}
	$S \leftarrow A(G)$\;	
	\For {$i \leftarrow 1 \ldots [\lceil \log_{3/2} n\rceil]$}{
		$W \leftarrow 9 \ln n / (3/2)^i$ \text{ sampled vertices uniformly at random from } $V(G)$\;
		$S_1 \leftarrow $\text{ projection of $S$ on columns of $W$}\;
		$S_2 \leftarrow $\text{ projection of $S$ on rows of $W$}\;
		$S \leftarrow S_1 \ttimes S_2$\; \label{sakht}
	}
	\Return $S$\;
	\caption{APSP for graphs of small weight}\label{algzw}
\end{algorithm}

Based on the arguments we explained above, \zwickall~\cite{zwick2002all} proved that Algorithm \ref{algzw} reports the APSP of a graph w.h.p. Now, we show that if the edge weights of a graph are small, we can implement Algorithm \ref{algzw} in a MapReduce setting in $O(1/\epsilon \log^2 n)$ MapReduce rounds. We assume that the memory of each machine is bounded by $O(n^\epsilon)$ and try to minimize the number of machines needed to implement Algorithm \ref{algzw}. Since the size of the input is $n^2$,  the number of machines in our algorithm is at least $O(n^{2-\epsilon})$ and thus all lines of Algorithm \ref{algzw} except Line \ref{sakht} can be trivially done in a single MapReduce round. Therefore, the only challenging part is to execute Line \ref{sakht} of Algorithm \ref{algzw}. In a step $i$, let $y = \log_n(3/2)^i$ and $x = 1-y$. Therefore, the sizes of the matrices $S_1$ and $S_2$ are $n \times \tildorder(n^x)$ and $\tildorder(n^x) \times n$, respectively. Moreover, we only care about the entries of the matrices in the range $[-n^y, n^y]$ and thus can assume w.l.g that all other values are  equal to $\infty$. Therefore, the range of the values in matrix multiplication is $O(n^y)$. Thus, we have two options for executing Line \ref{sakht} of the algorithm. We can either use the general $(\min, +)$ matrix multiplication algorithm of Theorem \ref{imbalancedplusmin} or the bounded range algorithm described in Theorem \ref{hadi2}. If the memory of the machines is $O(n^\epsilon)$ then the number of machines needed for the algorithm of Theorem \ref{imbalancedplusmin} is $\tildorder(n^{2(1-\epsilon/2)+(x-\epsilon/2)})$. The algorithm of Theorem \ref{hadi2} only works if $\epsilon \geq y$ and in that case the number of machines needed to compute $S_1 \ttimes S_2$ is $\tildorder(n^{\omega^*\langle 1,1,x\rangle(2-\epsilon+y)/2}) =  \tildorder(n^{\omega^*\langle 1,1,x\rangle(3-\epsilon-x)/2})$. However, we need to consider the following two limitations for the above algorithms:
\begin{itemize}
	\item Our algorithm for bounded $(\min, +)$ matrix multiplication (Theorem \ref{hadi2}) only works when $\epsilon \geq y$ and thus $x \geq 1-\epsilon$.
	\item Our algorithm for general $(\min, +)$ matrix multiplication (Theorem \ref{imbalancedplusmin}) only works when $x \geq \epsilon/2$. When $x < \epsilon/2$, we need to add additional zeros to $S_1$ and $S_2$ to make them of size $n \times n^{\epsilon/2}$ and $n^{\epsilon/2} \times n$ respectively. Therefore, in case $x < \epsilon/2$, the number of machines needed is $\tildorder(n^{2(1-\epsilon/2)})$.
\end{itemize}

Figure \ref{epsilons} illustrates the number of machines needed to take the $(\min, +)$ multiplication of the two matrices using each of the algorithms. Let 
$$\machines(\epsilon) = \begin{cases}
3-5/2\epsilon & \epsilon \leq 0.31924\\
\\
2(1-\epsilon/2) +&\\
{\frac{\sqrt{0.07\epsilon^2+1.66\epsilon + -0.43\epsilon+2.92 - 0.89}-0.26\epsilon-1.11}{0.53}-\epsilon/2}, & \text{otherwise}
\end{cases} $$ 
be a function defined over the domain $0 \leq \epsilon \leq 2$. We show in Theorem \ref{mainmain} that for any $0 \leq \epsilon \leq 2$, the APSP of a graph $G$ with $n$ vertices and edge weights in $\{-1,0,1\}$ can be computed with $\tildorder(n^{\machines(\epsilon)})$ machines with memory $O(n^\epsilon)$ in $O(1/\epsilon \log^2 n)$ MapReduce rounds.

\begin{theorem}\label{mainmain}
Let $0 \leq \epsilon \leq 2$ be a real number and $G$ be a graph with $n$ vertices whose edge weights are in $\{-1,0,1\}$. There exists a MapReduce algorithm to compute APSP of $G$  with $\tildorder(n^{\machines(\epsilon)})$ machines and memory $O(n^\epsilon)$ in $O(1/\epsilon \log^2 n)$ MapReduce rounds.
\end{theorem}
\begin{proof}
\zwickall~\cite{zwick2002all} proved that Algorithm \ref{algzw} compute APSP of a graph with high probability. Here, we show how this algorithm can be implemented in the MapReduce model. We assume the memory of each machine is $O(n^\epsilon)$ and the number of machines available is $\tildorder(n^{\machines(\epsilon)})$. Since $\machines(\epsilon) \geq 2-\epsilon$ for all $0 \leq \epsilon \leq 1$, we have enough memory to store a matrix of size $n \times n$. Therefore, each line of Algorithm \ref{algzw}, except Line \ref{sakht}, can be executed in a single MapReduce round. This makes a total of $O(\log n)$ MapReduce rounds. In order to compute $S_1 \ttimes S_2$, we use the better of the general algorithm for matrix multiplication and the algorithm for the bounded version. The former runs in $O(1/\epsilon)$ MapReduce rounds and the latter runs in $O(\log n)$ MapReduce rounds, so both algorithm need at most $O(1/\epsilon \log n)$ rounds. Therefore, the total number of MapReduce rounds in our implementation is $O(1/\epsilon \log ^2 n)$.

It only suffices to show that $\tildorder(n^{\machines(\epsilon)})$ machines are enough to compute $S_1 \ttimes S_2$ in each iteration. We assume that in every step $i$, $y = (3/2)^i$ is the range of the matrix values and $x=1-y$. As such, the sizes of the matrices to be multiplied in step $i$ are $n \times \tildorder(n^x)$ and $\tildorder(n^x) \times n$, respectively. In what follows, we determine for which $\epsilon$'s and which $x$'s each algorithm requires fewer machines. Note that if $x < \epsilon$, we cannot run the algorithm for bounded $(\min, +)$ matrix multiplication and can only determine the distance product via the general algorithm. As can be derived from the formula's and is apparent from Figure \ref{epsilons}, the bounded range algorithm needs fewer machines as $x$ increases. Likewise, one can see that the number of machines needed for the general algorithm increases as we increase $x$. Therefore, if at $x = 1-\epsilon$ the general algorithm needs more machines than the bounded range algorithm, this point is where the maximum number of machines is required for multiplying the matrices. In order to characterize such $\epsilon$'s we set $x = 1-\epsilon$ and thus the exponent of the number of machines for the bounded range multiplication algorithm is equal to
\begin{equation*}
\begin{split}
\omega^*\langle 1,1,x\rangle(3-\epsilon-x)/2 & = \omega^*\langle 1,1,x\rangle(3-\epsilon-(1-\epsilon)))/2 \\
& \leq \omega^*\langle 1,1,x\rangle\\
& \leq 1.8379 + 0.5347x \\
\end{split}
\end{equation*}
Furthermore, in this situation the exponent of the number of machines needed for the general algorithm can be formulated as
\begin{equation*}
\begin{split}
2(1-\epsilon/2)+(x-\epsilon/2) & = 2(1-(1-x)/2) + (x-(1-x)/2) \\
& = 2 (1/2+x/2) + (3/2x - 1/2)\\
& = (x+1) + (3/2x - 1/2) \\
& = 5/2x + 1/2
\end{split}
\end{equation*}
Notice that for $\epsilon \leq 0.31924$ we have $x \geq 0.68076$. Moreover $1.8379 + 0.5347x < 5/2x + 1/2$ holds for this range and therefore for $\epsilon \leq 0.31924$ the exponent of the number of machines necessary to compute $S_1 \ttimes S_2$ is $5/2x + 1/2 = 5/2(1-\epsilon)+1/2 = 3-5/2\epsilon$. Therefore $O(n^{\machines(\epsilon)})$ machines are enough to compute the multiplication for $\epsilon \leq 0.31924$.

Otherwise, the maximum number of machines we need is when the exponent of the number of necessary machines for both two algorithms are equal. Therefore, we have $$\omega^*\langle 1,1,x\rangle(3-\epsilon-x)/2  = 2(1-\epsilon/2)+(x-\epsilon/2).$$
Since $x \geq 1-\epsilon$ and $\epsilon \geq 0.31924$ we have $x \geq 0.696839$ and this implies $\omega^*\langle 1,1,x\rangle = 1.8379 + 0.5347x$. Therefore, in order for the terms to be equal we have 
\begin{equation}\label{bala}
\begin{split}
\omega^*\langle 1,1,x\rangle(3-\epsilon-x)/2 & = (1.8379 + 0.5347x) (3-\epsilon-x)/2\\
& = 2(1-\epsilon/2)+(x-\epsilon/2).
\end{split}
\end{equation}
Equation \eqref{bala} holds for $$x \simeq \frac{\sqrt{0.07\epsilon^2+1.66\epsilon + 0.53(-0.8379\epsilon+5.51) - 0.89}-0.26\epsilon-1.11}{0.53}$$ and in such cases we have $x \geq 1-\epsilon/2$ and we need 
$$\tildorder(n^{2(1-\epsilon/2) + ( \frac{\sqrt{0.07\epsilon^2+1.6\epsilon + 0.53(-0.83\epsilon+5.51) - 0.89}-0.26\epsilon-1.11}{0.53}-\epsilon/2)})$$
 machines. This is again equal to $O(n^{\machines(\epsilon)})$ and thus the proof is complete.
\end{proof}

Figure \ref{machines} compares $\machines(\epsilon)$ to the exponent of the general algorithm for finding APSP of unrestricted weighted graphs (Corollary \ref{apsp}).
Indeed the result of Theorem \ref{mainmain} holds for graphs with any integer edge weights in $[-R,R]$, with an additional factor $R$ appearing in the memory of the machines. By setting the number of machines equal to the memory of each machine we have:
\begin{corollary}[of Theorem \ref{mainmain}]\label{mainmain2}
	Let $G$ be a graph with $n$ vertices whose edge weights are in $\{-1,0,1\}$. There exists a MapReduce algorithm to compute APSP of $G$  with $\tildorder(n^{\starnumber})$ machines and memory $O(n^{\starnumber})$ in $O(\log^2 n)$ MapReduce rounds.
\end{corollary}

\section{Orthogonal Vectors}\label{orthogonalsection}
In the orthogonal vectors problem (\OV), we are given two lists of  vectors, each containing $n$ 0-1 vectors of size $O(\log n)$. Let the two lists be $A$ and $B$, and denote by $a_i$ ($b_i$) the $i$'th vector of $A$ ($B$). The goal of \OV\enspace is to find two vectors $a_i \in A$ and $b_j \in B$ such that $a_i . b_j = 0$. It is proven in ~\cite{impagliazzo1998problems} that \OV\enspace does not admit a truly subquadratic solution, unless \SETH\enspace fails.

\begin{theorem}[proven in ~\cite{impagliazzo1998problems}]\label{ortho1}
For any $\epsilon > 0$, there exists no $O(n^{2-\epsilon})$ time algorithm for solving \OV\enspace unless \SAT\enspace admits a $O(2^{(1-\epsilon')n})$ time solution for some $\epsilon' > 0$.
\end{theorem}

Note that an immediate consequence of Theorem \ref{ortho1} is that any MapReduce algorithm for \OV\enspace with a sublinear number of machines requires a super linear running time in order to solve the problem unless \SETH\enspace fails.  However, both the number of machines and the memory of each machine can be reduced to a sublinear number. In what follows, we show how one can efficiently solve \OV\enspace with $O(n^{2(1-\epsilon)})$ machines with memory $\tildorder(n^\epsilon)$.

The idea behind the algorithm is simple. We divide each of the lists into $n^{1-\epsilon}$ sublists of size $n^{\epsilon}$. Let these sublists be $A_1, A_2, \ldots, A_k$ and $B_1, B_2, \ldots, B_k$ where $k = n^{1-\epsilon}$. As such, we have $\cup A_i = A$ and $A_i \cap A_j = \emptyset$ for all $i \neq j$. Similarly, $\cup B_i = B$ and $B_i \cap B_j = \emptyset$ hold for all $i \neq j$. Now, one can reduce the problem to $k^2 = n^{2(1-\epsilon)}$ subproblem where every subproblem investigates the existence of orthogonal vectors  within a pair of sublists. These subtasks can be distributed between the machines and be solved in time $(n^{\epsilon})^2 = n^{2\epsilon}$ per machine. Finally, one can iterate over the solutions of all substasks and report any orthogonal vectors found by any machine.

\begin{theorem}\label{ortho}
	The orthogonal vectors problem can be solved with a MapReduce algorithm on $O(n^{2(1-\epsilon)})$ machines with memory $\tildorder(n^{\epsilon})$ in one MapReduce round. The total running time of the algorithm is $\tildorder(n^2)$.
\end{theorem}
\begin{proof}
As aforementioned, our algorithm runs in one MapReduce round. In the map stage, we divide the two lists into $n^{1-\epsilon}$ sublists of size $n^{\epsilon}$ and subsequently, in the reduce stage, each machine solves one of the $n^{2(1-\epsilon)}$ subtasks. Algorithm \ref{alg:orthogonal:mapper} describes the mappers for $\OV$. The lists are given to the mappers as a sequence of $\langle key; value\rangle$ pairs either in form of $\langle key; value\rangle = \langle (n,A,i); a_i\rangle$ to describe the elements of $A$ or $\langle key; value\rangle = \langle (n,B,i); b_i\rangle$ for the elements of $B$.

\begin{algorithm}

	\KwData{A key-value pair in the form of either  $\langle key; value\rangle = \langle (n,A,i); a_i\rangle$ or $\langle key; value\rangle = \langle (n,B,i); b_i\rangle$}
	\KwResult{A sequence of key-value pairs}
    $q \leftarrow \lceil n^{\epsilon} \rceil$\;	
    $c \leftarrow \lceil i/q \rceil$\;
	\If{$key$ is in the form of $(n,A,i)$}{
		\For {$j \in [ \lceil n/q \rceil ]$}{
			\textbf{output} $\langle (c,j); (A,i,a_i)\rangle$\;
		}
	}\Else{
		\For {$j \in [ \lceil n/q \rceil ]$}{
			\textbf{output} $\langle (j,c); (B,i,b_i)\rangle$\;
		}
	}
	
	\caption{\textsf{\OV\enspace Mapper}($A,  B$)}	\label{alg:orthogonal:mapper}
\end{algorithm}

Each mapper copies a pair of $\langle key; value\rangle$ into $n^{1-\epsilon}$ different key-value pairs, each to be fed to a separate machine. Each machine is identified with a pair $(i,j)$ where both $i$ and $j$ vary between $1$ and $\lceil n/q \rceil = O(n^{1-\epsilon})$. Therefore, each machine receives $q = O(n^{\epsilon})$ vectors of $A$ and $q$ vectors of $B$ in the reduce stage. Next, each machine receives the vectors and finds any orthogonal vectors via a loop over all pairs of vectors. Algorithm \ref{alg:orthogonal:reducer} shows how the reducers solve each subtask in the only round of the algorithm.

\begin{algorithm}

	\KwData{A key and a sequnce of values in the form of $\langle key; v_1, v_2, \ldots,v_l\rangle$}
	\KwResult{An output in the form of either $\langle key; \mathbf{false}\rangle$ or $\langle key; \mathbf{true}\rangle$}
	$answer \leftarrow \mathbf{false}$\;
	\For {$\alpha \in [l]$}{
		\If {$v_\alpha$ is in the form of $(A,i,a_i)$}{
			\For {$\beta \in [l]$}{
				\If {$v_\beta$ is in the form of $(B,j,a_j)$}{
					\If {$a_i.b_j = 0$}{
							$answer \leftarrow \mathbf{true}$\;
					}
				}
 		   }
		}
	}
	\textbf{output} $\langle key; answer\rangle$\;
	\caption{\textsf{\OV\enspace Reducer}($A,  B$)}	\label{alg:orthogonal:reducer}
\end{algorithm}
Notice that every pair of vectors $a_i \in A, b_j \in B$ is given to exactly one machine $(\lceil i/q\rceil, \lceil j/q\rceil)$ and thus any orthogonal pair will be detected by a machine. Thus, the reducers find an orthogonal pair if and only if the solution of the orthogonal vectors problem is positive. The number of machines used in this algorithm is $O(n^{2(1-\epsilon)})$ and the memory of each machine is $\tildorder(n^{\epsilon})$. Moreover, each machine has a running time of $\tildorder(n^{2\epsilon})$ and hence the total running time of the algorithm is $\tildorder(n^2)$.
\end{proof}

If one wishes to minimize the maximum of the number of machines and the memory of each machine, one can get a solution with $O(n^{2/3})$ machines and memory $\tildorder(n^{2/3})$ by setting $\epsilon = 2/3$.
\begin{corollary}[of Theorem \ref{ortho}]
	The orthogonal vectors problem can be solved with a MapReduce algorithm on $O(n^{2/3})$ machines with memory $\tildorder(n^{2/3})$ in a single MapReduce round. 
\end{corollary}
\onecolumn 
\section{Figures }

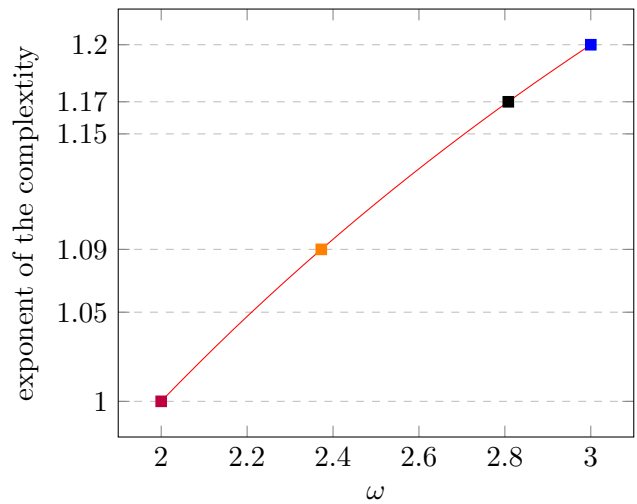
\begin{figure}[h]
\begin{center}
\begin{tikzpicture}
\begin{axis}[domain=2:3, title={The performance of our algorithm based on the value of $\omega$},     xlabel = $\omega$,
ylabel = {exponent of the complextity},
ytick={1,1.05,1.168,1.08523337983,1.15,1.2},
ymajorgrids=true,
grid style=dashed]
\addplot[color=red]{2*x/(x+2)};

\addplot[
color=blue,
mark=square*,
]
coordinates {
	(3,1.2)
};

\addplot[
color=black,
mark=square*,
]
coordinates {
	(2.808,1.168)
};

\addplot[
color=orange,
mark=square*,
]
coordinates {
	(2.3727,1.08523337983)
};

\addplot[
color=purple,
mark=square*,
]
coordinates {
	(2,1)
};


\end{axis}
\end{tikzpicture}
\end{center}
\caption{The complexity of the algorithm is illustrated for $\omega = 3$ (trivial algorithm), $\omega = 2.808$~\cite{strassen1969gaussian}, $\omega = 2.3728$~\cite{le2014powers}, and $\omega = 2$ (belief of the community).}
\label{omegas}\end{figure}

\begin{figure}[h]
\begin{center}
\begin{tikzpicture}
\begin{axis}[domain=0:1, title={The performance of imbalanced matrix multiplication based on the value of $x$},     xlabel = $x$,
ylabel = {exponent of the complextity},
ymajorgrids=true,
grid style=dashed
]
\addplot[color=red]{  (2*max(2,1.8379 + 0.5347*x))/ (max(2,1.8379 + 0.5347*x)+2) };


\end{axis}
\end{tikzpicture}
\end{center}
\caption{The exponent of the complexity of imbalanced matrix multiplication for $0 < x < 1$.}
\label{omegas2}\end{figure}
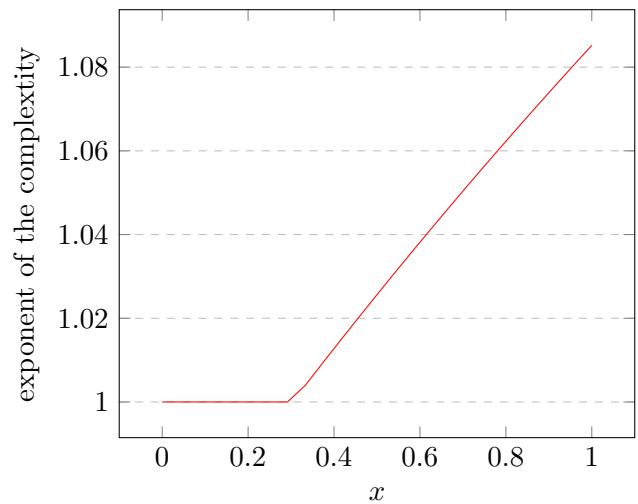

\clearpage

\begin{figure}[h]
\begin{center}
\hspace{-1cm}
\newcommand{\myvalue}{0.25}
\newcommand{\startdomain}{0.75}
\begin{subfigure}[b]{0.40\textwidth}
\begin{tikzpicture}
\begin{axis}[width=8cm,domain=0:1, title={$\epsilon = \myvalue$},     xlabel = $x$,
ylabel = {exponent of the number of machines},
ymajorgrids=true,
grid style=dashed
]
\addplot[color=red, domain=\startdomain:1]{  max(1.8379 + 0.5347*x, 2)*(3-\myvalue-x)/2 };
\addplot[color=blue]{  2*(1-\myvalue/2)+max(0,(x-\myvalue/2))};


\end{axis}
\end{tikzpicture}
\end{subfigure}
\hspace{1cm}
\renewcommand{\myvalue}{0.302}
\renewcommand{\startdomain}{0.68}
\begin{subfigure}[b]{0.40\textwidth}            
\begin{tikzpicture}
\begin{axis}[width=8cm,domain=0:1, title={$\epsilon = \myvalue$},     xlabel = $x$,
ylabel = {exponent of the number of machines},
ymajorgrids=true,
grid style=dashed
]
\addplot[color=red, domain=\startdomain:1]{  max(1.8379 + 0.5347*x, 2)*(3-\myvalue-x)/2 };
\addplot[color=blue]{  2*(1-\myvalue/2)+max(0,(x-\myvalue/2))};


\end{axis}
\end{tikzpicture}
\end{subfigure}

\hspace{-1cm}
\renewcommand{\myvalue}{0.5}
\renewcommand{\startdomain}{0.5}
\begin{subfigure}[b]{0.40\textwidth}
	\begin{tikzpicture}
	\begin{axis}[width=8cm,domain=0:1, title={$\epsilon = \myvalue$},     xlabel = $x$,
	ylabel = {exponent of the number of machines},
	ymajorgrids=true,
	grid style=dashed
	]
	\addplot[color=red, domain=\startdomain:1]{  max(1.8379 + 0.5347*x, 2)*(3-\myvalue-x)/2 };
	\addplot[color=blue]{  2*(1-\myvalue/2)+max(0,(x-\myvalue/2))};


	\end{axis}
	\end{tikzpicture}
\end{subfigure}
\hspace{1cm}
\renewcommand{\myvalue}{1}
\renewcommand{\startdomain}{0}
\begin{subfigure}[b]{0.40\textwidth}            
	\begin{tikzpicture}
	\begin{axis}[width=8cm,domain=0:1, title={$\epsilon = \myvalue$},     xlabel = $x$,
	ylabel = {exponent of the number of machines},
	ymajorgrids=true,
	grid style=dashed
	]
	\addplot[color=red, domain=\startdomain:1]{  max(1.8379 + 0.5347*x, 2)*(3-\myvalue-x)/2 };
	\addplot[color=blue]{  2*(1-\myvalue/2)+max(0,(x-\myvalue/2))};


	\end{axis}
	\end{tikzpicture}
\end{subfigure}

\hspace{-1cm}
\renewcommand{\myvalue}{1.5}
\renewcommand{\startdomain}{0}
\begin{subfigure}[b]{0.40\textwidth}
	\begin{tikzpicture}
	\begin{axis}[width=8cm,domain=0:1, title={$\epsilon = \myvalue$},     xlabel = $x$,
	ylabel = {exponent of the number of machines},
	ymajorgrids=true,
	grid style=dashed
	]
	\addplot[color=red, domain=\startdomain:1]{  max(1.8379 + 0.5347*x, 2)*(3-\myvalue-x)/2 };
	\addplot[color=blue]{  2*(1-\myvalue/2)+max(0,(x-\myvalue/2))};


	\end{axis}
	\end{tikzpicture}
\end{subfigure}
\hspace{1cm}
\renewcommand{\myvalue}{1.75}
\renewcommand{\startdomain}{0}
\begin{subfigure}[b]{0.40\textwidth}            
	\begin{tikzpicture}
	\begin{axis}[width=8cm,domain=0:1, title={$\epsilon = \myvalue$},     xlabel = $x$,
	ylabel = {exponent of the number of machines},
	ymajorgrids=true,
	grid style=dashed
	]
	\addplot[color=red, domain=\startdomain:1]{  max(1.8379 + 0.5347*x, 2)*(3-\myvalue-x)/2 };
	\addplot[color=blue]{  2*(1-\myvalue/2)+max(0,(x-\myvalue/2))};


	\end{axis}
	\end{tikzpicture}
\end{subfigure}

\end{center}
\caption{The exponent of the number of machines needed to run each of the algorithms for $\epsilon \in \{0.32,0.5,1,1.5\}$ is illustrated for $0 \leq x \leq 1$. Red lines show the performance of the bounded $(\min, +)$ multiplication based on a reduction from matrix multiplication and the blue lines show the performance of $(\min, +)$ multiplication for general inputs.}
\label{epsilons}\end{figure}
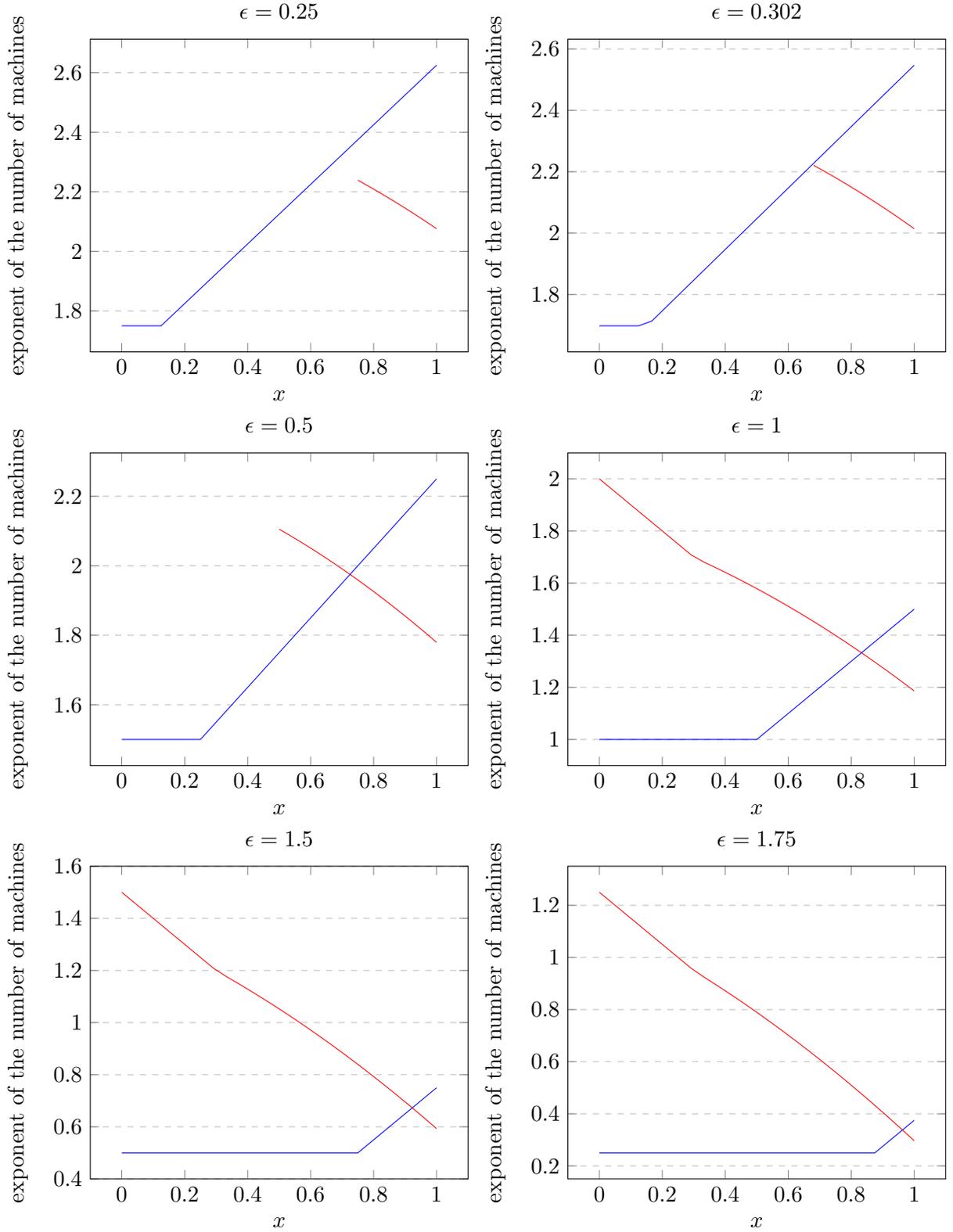

\clearpage

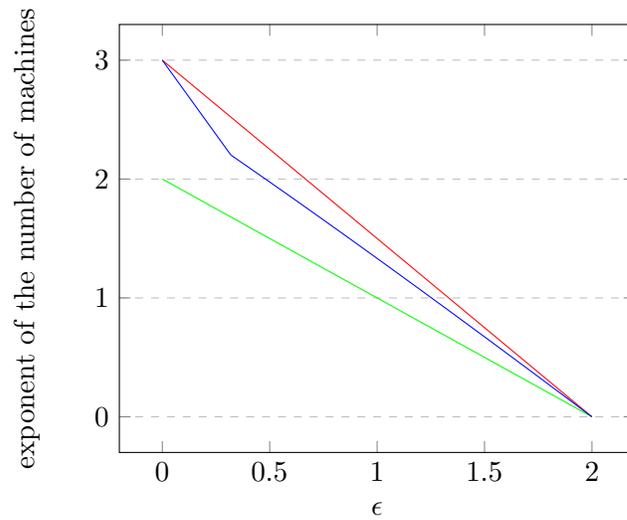
\begin{figure}[h]
\begin{center}
\begin{tikzpicture}
\begin{axis}[domain=0:2, title={The exponent of of the number of machines for APSP algorithms for memory $O(n^\epsilon)$},     xlabel = $\epsilon$,
ylabel = {exponent of the number of machines},
ymajorgrids=true,
grid style=dashed]
\addplot[color=red]{3*(1-x/2)};
\addplot[color=green]{2-x};
\addplot[color=blue,domain=0:0.31924]{3-5/2*x};
\addplot[color=blue,domain=0.31924:2]{2*(1-x/2) + ((sqrt(0.0714760225*x*x+1.66660643*x + 1.0694*(-0.8379*x+5.5137)/2 - 0.89133439)-0.26735*x-1.1169)/0.5347-x/2)};


\end{axis}
\end{tikzpicture}
\end{center}
\caption{Blue polyline shows the exponent of the number of machines needed to find the APSP matrix of a graph with small weights and red polyline shows this for general unrestricted graphs. We assume that the memory of each machine is $O(n^\epsilon)$. The green line shows $2-\epsilon$ which is a lower bound on the exponent of the number of machines for any MapReduce algorithm for APSP that has $O(n^\epsilon)$ memory on each machine.}
\label{machines}\end{figure}

\clearpage

\end{document}